\documentclass[a4paper,11pt,final]{elsarticle}

\makeatletter
\def\ps@pprintTitle{%
 \let\@oddhead\@empty
 \let\@evenhead\@empty
 \def\@oddfoot{\centerline{\thepage}}
 \let\@evenfoot\@oddfoot}
\makeatother

\usepackage[a4paper, left=.8in, right=.8in, top=.9in, bottom=.9in]{geometry} 

\usepackage{times}

\usepackage[colorlinks=true,citecolor=blue,linkcolor=blue,urlcolor=blue]{hyperref}

\usepackage[utf8]{inputenc}
\usepackage{graphicx}
\usepackage{amsmath}
\usepackage{amssymb}

\usepackage{braket}
\usepackage{float}

\usepackage[capitalise]{cleveref}
\crefname{equation}{Eq.\!}{Eqs.\!}
\crefname{figure}{Fig.\!}{Figs.\!}
\crefname{chapter}{Chap.\!}{Chaps.\!}
\crefname{section}{Sec.\!}{Secs.\!}
\crefname{appendix}{}{}

\renewcommand{\P}{\mathcal{{P}}}
\newcommand{\DD}{\mathbb{D}}
\newcommand{\RR}{\mathbb{R}}
\newcommand{\CC}{\mathbb{C}}

\begin{document}

\title{Transfer efficiency enhancement and eigenstate properties in \\ locally symmetric disordered finite chains}

\author[1]{C.~V. Morfonios}

\author[1]{M. R\"ontgen}

\author[2]{F.~K. Diakonos}

\author[1,3]{P. Schmelcher}

\address[1]{Zentrum f\"ur Optische Quantentechnologien, Universit\"{a}t Hamburg, Luruper Chaussee 149, 22761 Hamburg, Germany}
\address[2]{Department of Physics, University of Athens, Panepistimiopolis, 15771 Athens, Greece}
\address[3]{The Hamburg Centre for Ultrafast Imaging, Universit\"{a}t Hamburg, 22761 Hamburg, Germany 
 \begin{center}
 \normalfont
 Journal reference: \href{https://doi.org/10.1016/j.aop.2020.168163} {{Annals of Physics}\ \textbf{418}, {168163} ({2020}) }
\vspace{-7ex}
 \end{center}
}

\date{}

\begin{abstract}
The impact of local reflection symmetry on wave localization and transport within finite disordered chains is investigated.
Local symmetries thereby play the role of a spatial correlation of variable range in the finite system.
We find that, on ensemble average, the chain eigenstates become more fragmented spatially for intermediate average symmetry domain sizes, depending on the degree of disorder.
This is caused by the partial formation of states with approximate local parity confined within fictitious, disorder-induced double wells and perturbed by the coupling to adjacent domains.
The dynamical evolution of wave-packets shows that the average site-resolved transfer efficiency is enhanced between regions connected by local symmetry.
The transfer may further be drastically amplified in the presence of spatial overlap between the symmetry domains, and in particular when global and local symmetry coexist.
Applicable to generic discrete models for matter and light waves, our work provides a perspective to understand and exploit the impact of local order at multiple scales in complex systems.
\end{abstract}


\maketitle

\section{Introduction}

Since the pioneering theoretical work by Anderson \cite{Anderson1958_PR_109_1492_AbsenceDiffusionCertainRandom} it has been shown that, under many circumstances, spatial disorder in a medium leads to the exponential localization of waves due to multiple destructive interference. 
This behavior in turn suppresses the transport of an initial wave excitation through a disordered sample between remote sites. 
Initially explored for electrons in models of solids \cite{Anderson1958_PR_109_1492_AbsenceDiffusionCertainRandom,Thouless1974_PR_13_93_ElectronsDisorderedSystemsTheory}, disorder-induced localization has more recently also been demonstrated and intensively studied for light waves \cite{Schwartz2007_N_446_52_TransportAndersonLocalizationDisordered,Segev2013_NP_7_197_AndersonLocalizationLight,Bertolotti2005_PRL_94_113903_OpticalNecklaceStatesAnderson,Sheinfux2016_NC_7_12927_InterplayEvanescenceDisorderDeep} as well as for matter waves in optical lattices \cite{Billy2008_N_453_891_DirectObservationAndersonLocalization,Roati2008_N_453_895_AndersonLocalizationNoninteractingBoseEinstein,Pasek2017_PRL_118_170403_AndersonLocalizationUltracoldAtomsa}.
Typically manifest in one-dimensional (1d) discrete lattice models with random onsite potential (diagonal disorder) or inter-site hoppings (off-diagonal disorder) \cite{Kondakci2015_NP_11_930_PhotonicThermalizationGapDisordered,Kondakci2016_OO_3_477_SubthermalSuperthermalLightStatistics}, localization also occurs for structural disorder in systems with binary constituents \cite{Bertolotti2005_PRL_94_113903_OpticalNecklaceStatesAnderson,Naether2013_NJP_15_13045_ExperimentalObservationSuperdiffusiveTransport} and may further be induced in the bulk of setups with disordered boundary \cite{Naether2012_OL_37_485_AndersonLocalizationPeriodicPhotonic}.

The presence of spatial correlations between the constituents of an otherwise disordered medium generally causes wave excitations to be less localized and enhances transport, with the detailed system response depending on the type of correlated disorder \cite{Izrailev2012_PR_512_125_AnomalousLocalizationLowdimensionalSystems}.
Correlation can be short-ranged \cite{Titov2005_PRL_95_126602_NonuniversalityAndersonLocalizationShortRange,Sales2012_PELSaN_45_97_NumericalStudyOneelectronDynamics}, 
in the form of ordered clustered elements such as dimers \cite{Naether2013_NJP_15_13045_ExperimentalObservationSuperdiffusiveTransport,Dunlap1990_PRL_65_88_AbsenceLocalizationRandomdimerModel,Lavarda1994_PRL_73_1267_ResonantScatteringNonsymmetricDimers,Sanchez1994_PRB_49_147_SuppressionLocalizationKronigPenneyModels}, trimers \cite{Datta1994_JPCM_6_4465_AbsenceLocalizationOnedimensionalDisordered}, or polymers \cite{Izrailev1999_PRL_82_4062_LocalizationMobilityEdgeOneDimensional,Lopez-Gonzalez2016_PRE_93_32205_TransportLocalizedExtendedExcitations}, 
it can have long-range character \cite{Izrailev2012_PR_512_125_AnomalousLocalizationLowdimensionalSystems,deMoura1998_PRL_81_3735_Delocalization1DAndersonModel,Cheraghchi2005_PRB_72_174207_LocalizationdelocalizationTransitionOneOnedimensional}, 
while also mixed short- and long-range correlations \cite{Guo2011_PRB_83_245108_SuppressionLocalizationTwolegLadder} 
as well as subsystem disorder \cite{Moura2011_JPCM_23_135303_ResonantStatesWavepacketSuperdiffusion} have been explored.
Further, delocalization may be facilitated by correlations between onsite elements alone \cite{Carvalho2011_JPCM_23_175304_LocalizationTwochannelModelCrosscorrelated}, between hoppings \cite{Cheraghchi2005_PRB_72_174207_LocalizationdelocalizationTransitionOneOnedimensional}, or between onsite and hopping elements \cite{Dunlap1990_PRL_65_88_AbsenceLocalizationRandomdimerModel}.

\begin{figure*}[t]
\center
\includegraphics[width=\textwidth]{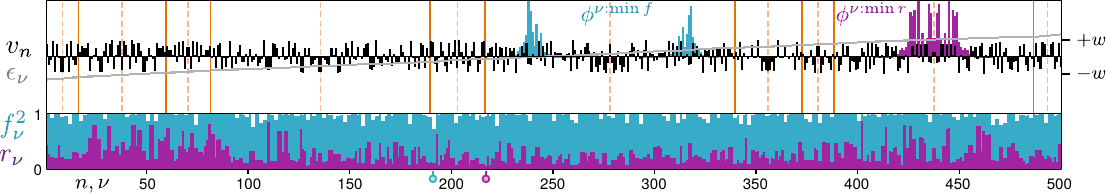}
\caption{
\textbf{Top}: Locally reflection-symmetric disordered (LRD) chain of $N = 500$ sites for disorder strength $w = 3h$, with initially random onsite elements $v_n \in [-w,w]$ symmetrized around local symmetry centers $c_{d= 1,2, \dots,D}$ (vertical dashed lines) of $D = 10$ domains attached at random interfaces (vertical solid lines), and corresponding Hamiltonian eigenvalues $\epsilon_\nu$, for uniform hopping $h$ (set to $0.01$).
\textbf{Bottom}: Inverse participation ratio (IPR) $r_\nu$  and cumulative Friedel sum (CFS) $f_\nu$ for increasing eigenstate index $\nu \in [1,N]$.
The purple (blue) circle indicates the eigenmode of minimal $r_\nu$ ($f_\nu$), with correspondingly colored modulus $|\phi^\nu_n|$ plotted in the top panel.
}
\label{fig:setupExample}
\end{figure*}

In the meanwhile vast literature on wave localization, the impact of \textit{spatial symmetry} has largely been used on the small scale of lattice constituents in resonant conditions of transport.
On the other hand, a series of recent studies shows that the presence of global \textit{centrosymmetry} \cite{Zech2014_NJP_16_55002_CentrosymmetryEnhancesQuantumTransport}---equivalent to reflection symmetry for 1d systems---in an otherwise disordered finite system may lead to a prevalence of extended states and thereby enhance wave transfer between symmetry related sites \cite{Zech2014_NJP_16_55002_CentrosymmetryEnhancesQuantumTransport,Walschaers2013_PRL_111_180601_OptimallyDesignedQuantumTransport,Walschaers2015_PRE_91_42137_StatisticalTheoryDesignedQuantum,Walschaers2016_ARCMP_7_223_QuantumTransportDisorderedNoisy}.
This principle applies to general multi-dimensional networks of interconnected nodes, even in the presence of many-body interactions \cite{Ortega2016_PRE_94_42102_EfficientQuantumTransportDisordered,Ortega2018_PRE_98_12141_RobustnessOptimalTransportDisordered}.
The effect is closely connected to the definite parity of the system eigenmodes which may result in correspondly more extendedstates on configuration average \cite{Walschaers2015_PRE_91_42137_StatisticalTheoryDesignedQuantum}.
Exploiting this property, global reflection symmetry has also been proposed as a generator of tunneling in 1d disordered potentials with applications to secure communication in classical circuits \cite{Diez2008_PRB_78_35118_SymmetryinducedTunnelingOnedimensionalDisordered}.
The behavior of such globally symmetric systems in terms of optimal transfer efficiency is subject to further design conditions \cite{Zech2014_NJP_16_55002_CentrosymmetryEnhancesQuantumTransport,Walschaers2013_PRL_111_180601_OptimallyDesignedQuantumTransport}, but generally demonstrates the crucial role of symmetry coexisting with disorder. 

In fact, global symmetry is seldom exactly fulfilled. 
Continuous symmetry measures \cite{Zabrodsky1992_JACS_114_7843_ContinuousSymmetryMeasures} and symmetry operation measures \cite{Pinsky2008_JCC_29_190_SymmetryOperationMeasures} have been proposed to describe deviation from exact symmetry. 
A different paradigm of global symmetry breaking is the case of exact but \textit{local} symmetries, that is, symmetries which are fulfilled in a restricted subdomain of a composite system.
A recently developed theoretical framework addresses such local symmetries in terms of symmetry-adapted non-local currents \cite{Kalozoumis2013_PRA_87_32113_LocalSymmetriesOnedimensionalQuantum} governed by generalized non-local continuity equations \cite{Morfonios2017_AoP_385_623_NonlocalDiscreteContinuityInvariant,Rontgen2017_AoP_380_135_NonlocalCurrentsStructureEigenstates}. 
Their stationary versions reveal the presence of 1d local symmetries in generic wave-mechanical systems including non-Hermitian \cite{Morfonios2017_AoP_385_623_NonlocalDiscreteContinuityInvariant,Kalozoumis2016_PRA_93_63831_PTsymmetryBreakingWaveguidesCompeting} or even driven
\cite{Morfonios2017_AoP_385_623_NonlocalDiscreteContinuityInvariant,Wulf2016_PRE_93_52215_ExposingLocalSymmetriesDistorted} setups. 
In particular, they enable amplitude mappings which generalize the parity and Bloch theorems to the case of local symmetry \cite{Kalozoumis2014_PRL_113_50403_InvariantsBrokenDiscreteSymmetries}, and can be used to classify perfectly transmitting states \cite{Morfonios2017_AoP_385_623_NonlocalDiscreteContinuityInvariant,Kalozoumis2013_PRA_88_33857_LocalSymmetriesPerfectTransmission}.
A well-known class of systems featuring abundant local symmetries is that of 1d binary deterministic aperiodic structures (see e.\,g. Ref.\,\cite{Morfonios2014_ND_78_71_LocalSymmetryDynamicsOnedimensional} and references therein), where their combinatorial properties have been studied in terms of the so-called ``palindrome complexity'' \cite{Allouche2003_TCS_292_9_PalindromeComplexity}.
Local symmetries may generally also be found ``hidden'' \cite{Wochner2009_P_106_11511_XrayCrossCorrelationAnalysis} in amorphous and disordered systems \cite{Altarelli2010_PRB_82_104207_XrayCrosscorrelationAnalysisLocal} where their presence may affect order-disorder transitions \cite{Coupier2005_PRE_71_46105_LocalSymmetriesOrderdisorderTransitions}, or exist ``concurrently'' in interplay with global symmetries within molecules \cite{Echeverria2011_CEJ_17_359_ConcurrentSymmetriesInterplayLocal}.
Ultimately, any system with global symmetry which is coupled to an environment can be considered locally symmetric.
In view of the manifest role of global symmetry, this abundance of local symmetries raises the question of their impact on localization and transfer when multiply present at different locations and scales in disordered systems.

We here view local reflection symmetry, defined within different spatial subparts of a system, as a particular type of correlation of fixed or variable range in an otherwise disordered finite system. 
To study the effect solely of local symmetry in a simple setting, we consider finite 1d tight-binding chains with disordered onsite potentials which are mirror-symmetrized within adjacent or overlapping spatial domains of random or uniform size.
The localization properties of the eigenstates of such locally reflection-symmetric disordered (LRD) chains are then studied numerically for varying disorder strength and symmetry domain sizes (for brevity, from now on ``symmetry'' will refer exclusively to reflection symmetry, unless otherwise stated).
Apart from the widely used inverse participation ratio \cite{Rodriguez2006_JPAMG_39_14303_OnedimensionalModelsDisorderedQuantum}, we utilize a recently proposed \cite{Gong2016_PLA_380_59_MeasureLocalizationPropertiesOnedimensional} alternative localization measure which reflects the fragmentation of states induced by the local symmetries. 
An intricate interplay between the short-range localization and and long-range fragmentation properties is observed. 
It indicates an overall change to more fragmented states, within the ensemble average, for intermediate degree of local symmetry, with the uncorrelated case recovered in the limit of small symmetry domains.
This behavior is analyzed by combining the notion of fictitious disorder-induced tunneling barriers with the concept of symmetrization of eigenstates into symmetry domains, in turn explained within a local resonant scattering picture.
A crucial ingredient is here the concept of approximate \textit{local parity} \cite{Kalozoumis2013_PRA_87_32113_LocalSymmetriesOnedimensionalQuantum,Morfonios2017_AoP_385_623_NonlocalDiscreteContinuityInvariant,Kalozoumis2014_PRL_113_50403_InvariantsBrokenDiscreteSymmetries} of localized eigenstates within symmetry domains perturbed by the coupling to adjacent domains.
Notably, the purpose here is to investigate the effect of the local symmetry correlations on the properties of \textit{finite} LRD chains, which do not serve to approximate the large chain limit; they are simply chosen large enough to vary the number of symmetry domains and to perform eigenstate statistics.
Ultimately, we explore the impact of local symmetry on the diffusion of time-evolved wave-packets, by computing statistical distributions of the site-resolved transfer efficiency upon a single site excitation in LRD chains with few symmetry domains.
We here show that local symmetry may significantly enhance the transfer depending on the domain configurations.
A drastic increase in transfer is shown to occur when symmetry domains overlap with each other.
In particular, the transfer enhancement induced by global symmetry can be further increased considerably when local symmetry is present simultaneously at smaller scales.

The paper is organized as follows. 
In \cref{sec:eigenstateLocalization} we first define the considered LRD chain setups and provide the analysis tools used to distinguish localization from fragmentation (\cref{sec:IPRvsCFS}).
We then classify the types of eigenstate profiles present in the LRD chains (\cref{sec:symmetrization}) which are employed to explain the distribution of the computed localization measures for varying disorder and symmetry (\cref{sec:statAnalysis}).
In \cref{sec:dynamics} we investigate wave-packet dynamics in LRD chains, demonstrating the enhancement of transfer efficiency via local symmetry (\cref{sec:transfEff}) and its further increase in the presence of symmetry domain overlaps (\cref{sec:transfEffOverlap}).
Section \ref{sec:conclusions} concludes our investigations.
\cref{app:localScattering} explains the typical ``symmetrization'' of eigenstates into symmetry domains and \cref{app:fictitiousBarriers} provides a mapping of such eigenstates to ``fictitious'' double wells.
\cref{app:eigenstateSymmetryStat} corroborates the localization features with eigenstate symmetrization statistics, \cref{app:noVersusGlobalSymmetry} analyzes the effect of global symmetry and its statistics for larger chains, and \cref{app:spectralStat} focuses on the spectral statistics of LRD chains.

\section{Eigenstate fragmentation in locally reflection-symmetric disordered (LRD) lattices \label{sec:eigenstateLocalization}}

We consider a generic 1d chain of $N$ sites with uniform real next-neighbor hopping $h$ described by the single-particle Hamiltonian 
\begin{equation} \label{eq:hamiltonian}
 H = \sum_n v_n \ket{n}\bra{n} + \sum_{|m-n| = 1} h \ket{m}\bra{n}
\end{equation}
where $v_n$ is the onsite potential of site $n$ with single site orbital $\ket{n}$.
The onsite potential values are taken from a uniform random distribution with magnitude up to a disorder strength parameter $w$, that is, $v_n \in [-w,+w]$.
The potential array is then locally symmetrized within $D$ adjacent domains $\DD_d$ ($d = 1,2,\dots,D$) of sizes (that is, number of contained sites) $N_d$, starting from the left.
The resulting chain is thus symmetric under the action of $D$ local reflection transformations $\P_{c_d;N_d} \equiv \P_{\DD_d}$, each of which performs a permutation of sites only within domain $\DD_d$ about its center $c_d$ and acts as the identity on the rest of the chain \cite{Kalozoumis2013_PRA_87_32113_LocalSymmetriesOnedimensionalQuantum,Morfonios2017_AoP_385_623_NonlocalDiscreteContinuityInvariant},
\begin{equation} \label{eq:localSymmetry}
\P_{\DD_d}: n \to 
\begin{cases}
2c_d - n, \quad &n \in \DD_d \\
n, \quad &n \notin \DD_d 
\end{cases}
\end{equation}
In other words, our chain can be constructed by concatenating $D$ reflection symmetric subdomains.
In order to study the impact solely of the presence of local symmetry, to begin with the $N_d$ are also taken at random from a uniform distribution, obeying $\sum_d N_d = N$ with $N_d \geqslant 1$.
A correlation of spatially variable range $N_d$ is therefore induced into the otherwise disordered chain.
Such a configuration is shown in \cref{fig:setupExample} (top panel).

In the following we investigate the localization properties of the eigenvectors $\ket{\phi^\nu} = \sum_n \phi^\nu_n \ket{n}$ of $H$, given by 
\begin{equation} \label{eq:evp}
 H \ket{\phi^\nu} = \epsilon_\nu \ket{\phi^\nu},
\end{equation}
with eigenvalues $\epsilon_\nu$.
The spatial profiles of the squared eigenmode norms $\rho^\nu_n = |\phi^\nu_n|^2$ are unaffected by the sign of the hopping $h$ which induces a relative $\pi/2$ phase flip between adjacent sites \cite{Lyra2015_E_109_47001_DualLandscapesAndersonLocalization}.
We have here chosen $h > 0$, modeling e.\,g. the evanescent coupling between photonic waveguides, while the choice $h<0$ would correspond to e.\,g. the kinetic energy of non-interacting electrons on a tight-binding lattice.

\subsection{Localization versus fragmentation of states \label{sec:IPRvsCFS}}

A convenient and widely used single-parameter indicator of the grade of localization of a wavefunction is the inverse participation ratio (IPR) defined by \cite{Rodriguez2006_JPAMG_39_14303_OnedimensionalModelsDisorderedQuantum} 
\begin{equation} \label{eq:ipr}
 r = \sum_{n=1}^N \rho_n^2 ~~\in [N^{-1},1]
\end{equation}
for a normalized state $\ket{\psi}$ of squared modulus $\rho_n = |\psi_n|^2$ (with $\sum_{n=1}^N \rho_n = 1$).
The IPR takes on its maximal value $r = 1$ in the limit of a state localized on a single site $m$, $\rho_n = \delta_{mn}$, and its minimal value $r = 1/N$ for a state uniformly extended over the chain, $\rho_n = 1/N$.
As desired for a localization measure, the IPR does not depend on the position at which a state is localized within a disordered system.
At the same time, however, it is also largely insensitive to the spatial state profiles \cite{Gong2016_PLA_380_59_MeasureLocalizationPropertiesOnedimensional}, which in general do affect the static properties and dynamical response of the system.
An alternative localization measure, proposed very recently in Ref.\,\cite{Gong2016_PLA_380_59_MeasureLocalizationPropertiesOnedimensional} and inspired by the Friedel sum rule \cite{Langreth1966_PR_150_516_FriedelSumRuleAndersons}, reflects more details of the spatial profile $\rho_m$ by using its cumulative sum $P_n = \sum_{m=1}^n \rho_m$ up to site $n$.
We slightly redefine (see comment below) the measure here as
\begin{equation} \label{eq:cfs}
 f = \frac{1}{2N} \left|\sum_{n=1}^N \left(e^{2\pi i P_n} + 1\right)  \right| ~~\in [N^{-1},1],
\end{equation}
which we will refer to as the ``cumulative Friedel sum'' (CFS) of a given state.
Again, larger (smaller) CFS indicates a more (less) localized state, though now taking into account its total spatial extent instead of only its site participation, as described in the following.

The IPR and CFS distributions among the eigenmodes of a single LRD chain configuration are shown in the bottom panel of \cref{fig:setupExample}.
An impression of the difference between IPR and CFS in indicating localization properties is provided by the eigenstates $\phi^{\nu:\min r}$ and $\phi^{\nu:\min f}$ in \cref{fig:setupExample} having minimal $r$ and $f$, respectively, for the example setup.
With a similar density contribution (comparable amplitudes at similar number of sites), the states have almost the same IPR, which thus does not distinguish them.
In contrast, the drastically smaller CFS of $\phi^{\nu:\min f}$ indicates its extended profile:
The envelope consists of two individual maxima which are more peaked than in $\phi^{\nu:\min r}$, but lie farther apart, thus yielding an increased total extent.
For (normalized) states with local maxima, the CFS can thus be seen to indicate the degree of spatial ``fragmentation'' , that is, how remote from each other the amplitude maxima are located.
As an example, if we consider a (virtual) normalized state consisting of two single-site peaks at spacing $s$ and zero elsewhere in an $N$-site chain, then the CFS decreases monotonously from $f = 1$ at $s = 0$ (one single-site peak) to $f = 1/N$ at $s = N-1$ (one peak at each end of the chain) 
\footnote{For comparison, the measure $|\sum_{n=1}^N e^{2\pi i P_n}/N|$ in Ref.\,\cite{Gong2016_PLA_380_59_MeasureLocalizationPropertiesOnedimensional} would tend to $0$ for spacing $s \to N/2$ for large $N$ and then rise towards $1$ again for $s \to N$. This would make its value ambiguous for (locally) symmetric $\rho_n$ with large spacings. We have therefore redefined the measure into $f$ in \cref{eq:cfs} to better serve the present study.
}.
If each of the two peaks has a symmetric profile of common finite width (in the form of, e.\,g., a Gaussian or a rectangular step), then $f$ is independent of this width.
Thus, the CFS \textit{complements} the IPR, as a localization measure which is sensitive to the spacing of peaks in a wavefunction but relatively insensitive to the width of the peaks themselves (except for single peaks, that is, of non-fragmented states).

Before we present the statistical behavior of the IPR and CFS in \cref{sec:statAnalysis}, we next provide an intuitive interpretational tool where the symmetry domains effectively behave like double wells perturbed by the coupling to adjacent domains.

\subsection{Eigenstate symmetrization \label{sec:symmetrization}}

The qualitative distinction between the IPR and CFS in the present context of LRD chains is closely linked to the fact that, as explained in \cref{app:localScattering}, the eigenstates generally tend to ``symmetrize'' into the symmetric chain domains.
By this we mean that, for a sufficiently localized eigenvector $\ket{\phi^\nu}$ in a LRD finite chain, the density will have the \textit{approximate symmetry} $\rho_n \approx \rho_{2c_d - n}$ about the center $c_d$ of some domain $\DD_d \owns n$, while approximately \textit{vanishing outside} of it, $\rho_{n \notin \DD_d} \approx 0$.
Examples of this are states $\phi^{\nu:\min r}$ and $\phi^{\nu:\min f}$ already seen in \cref{fig:setupExample}.
In other words, the states tend to become approximate \textit{local parity eigenstates} \cite{Morfonios2017_AoP_385_623_NonlocalDiscreteContinuityInvariant} of local reflections $\P_{\DD_d}$ as defined in \cref{eq:localSymmetry} \footnote{Note that local symmetry in $\rho_n$ is not automatically fulfilled by a symmetry in the total Hamiltonian. Indeed, despite the geometric symmetry of the chain under the $\P_{\DD_d}$, those operations do not commute with $H$ since they do not preserve the connectivity of the chain \cite{Morfonios2017_AoP_385_623_NonlocalDiscreteContinuityInvariant}: Domain end sites are coupled to different sites upon transformation.}.

\begin{figure}[t]
\center
\includegraphics[width=.6\columnwidth]{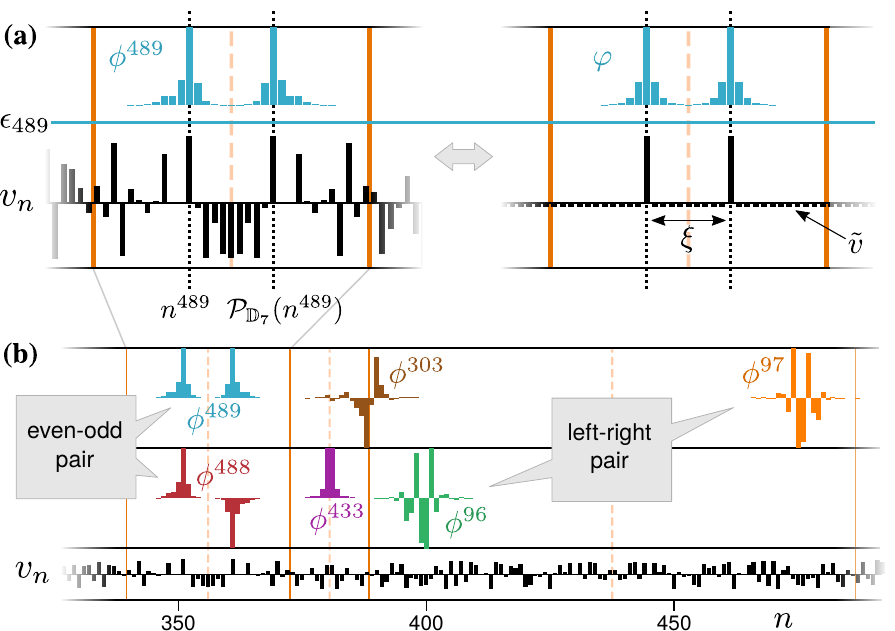}
\caption{
\textbf{(a)} Focus on (\textit{left}) the eigenstate $\phi^{489}$ localized symmetrically in domain $\DD_7$ of the chain in \cref{fig:setupExample}, with potential $v_n$ shown in arbitrary units, and (\textit{right}) the corresponding ``fictitious'' state $\varphi$ decaying exponentially away from the maximum positions $n_\nu,\P_{\DD_d}(n_\nu)$ within a constant potential $\tilde{v}$ (see \cref{eq:fictitiousPotential}), with the potential at $n_\nu,\P_{\DD_d}(n_\nu)$ tuned so that $\varphi$'s energy equals $\epsilon_{489}$ (indicated by dotted line).
The disordered symmetry domain is thus associated to a fictitious double well with barrier width $\xi$ (see \cref{eq:fictitiousWidth}) for this particular (symmetrized) eigenstate.
\textbf{(b)} State types in the LRD chain of \cref{fig:setupExample}: Even-odd state pair (with indices $\nu = 488,489$) in domain $\DD_7$, left-right state pair ($\nu = 96,97$) in $\DD_9$, single even state ($\nu = 433$) in $\DD_8$, and asymmetric state ($\nu = 303$) localized at the boundary between $\DD_8$ and $\DD_9$. 
States are normalized to maximum modulus and plotted offset for visibility.
}
\label{fig:selectedStates}
\end{figure}

Further, as explained in \cref{app:fictitiousBarriers}, each such locally symmetrized eigenstate can be mapped to a \textit{``fictitious double well''} with constant inter-well barrier strength $\tilde{v} = \tilde{v}^\nu$ and of width $\xi = \xi^\nu_d$ (corresponding to state $\ket{\phi^\nu}$ symmetrized into domain $\DD_d$) given by the spacing between the state's maxima; see \cref{eq:fictitiousPotential,eq:fictitiousWidth}, respectively. 
This mapping is visualized in \cref{fig:selectedStates}\,(a) for a selected state localized in domain $\DD_7$ of the setup in \cref{fig:setupExample}.
Notably, for the eigenstate symmetrization to occur in a domain $\DD_d$, the short-range localization length $\ell$ of the state (see \cref{app:fictitiousBarriers}) should be significantly smaller than that domain's size $N_d$.

In terms of local symmetry, the eigenstates of a finite LRD chain will in general be of one of the following types, with examples shown in \cref{fig:selectedStates}\,(b):

(i) ``Even-odd'' ($eo$) pair: two quasidegenerate states of approximate even and odd local parity, resembling the energy-split states of an isolated symmetric double well, with approximately the same density profile (see $\phi^{488},\phi^{489}$); 

(ii) ``Left-right'' ($lr$) pair: two quasidegenerate states localized in the left and right half of a symmetry domain each, resembling the states of an isolated well and its mirror image, each with spatial state profile being approximately the mirror image---under $\P_{\DD_d}$---of the other (see $\phi^{96},\phi^{97}$);

(iii) Single states of approximate local (even or odd) parity (see $\phi^{433}$), in cases where the above-mentioned fictitious barrier width $\xi$ is of the order of the short-range localization length $\ell$ (see \cref{app:fictitiousBarriers});

(iv) Single asymmetric states localized around the boundary between two symmetry domains for sufficiently strong disorder (see $\phi^{303}$) or extended over multiple domains for very weak disorder, sharing none of the above properties.

We emphasize here that the coupling of the symmetry domain boundaries to the surroundings (adjacent domains) acts as a \textit{perturbation} on the local parity of domain-localized eigenstates.
This perturbation increases with the overlap of those states with the domain boundaries, and depending on the fictitious inter-well barrier (see \cref{app:fictitiousBarriers}), they may (like $eo$ pairs) or may not (like $lr$ pairs) have approximate local parity with respect to $\P_{\DD_d}$.
Indeed, $lr$ pairs can be seen as originating from $eo$ pair states which are practically degenerate due to vanishingly small inter-well coupling (large $\xi^\nu_d$ and/or $\tilde{v}^\nu$ of the fictitious barrier) and combine linearly into left- and right-localized states under the boundary perturbation \footnote{Variants of $lr$ pair states may thereby further occur which are mainly localized on one domain half but have a small amplitude also on the other half, resulting from the combination of $eo$ pair states with slightly asymmetric (again due to broken parity) densities.}.
In other words, a stronger disorder-induced fictitious double-well barrier assists the local parity breaking caused by the coupling of the domain to its environment.

The IPR and CFS distribution among the eigenstates of a given LRD chain will highly depend on the occurrence of $eo$ and $lr$ pairs.
For $eo$ pair states, a localization peak at some position, denoted $n_\nu$, within a symmetry domain $\DD_d$ imposes the same localization peak at $\P_{\DD_d}(n_\nu)$.
This yields relatively small IPR and CFS values, each with a double multiplicity (since the pair states have almost the same density)---as evident, e.\,g., from pairs of equal consecutive $r_\nu$- or $f_\nu$-bars in \cref{fig:setupExample}.
The CFS will additionally decrease with the distance $\xi^\nu_d $ between the two density peaks which represents the degree of the state's fragmentation mentioned in \cref{sec:IPRvsCFS}.
In contrast, $lr$ state pairs contribute with relatively high IPR and CFS values (now without fragmentation present), again with double multiplicity.

The relative frequency of $eo$ and $lr$ pairs will depend on the average fictitious double-well barriers emerging among the different domains.
For a given moderate disorder strength, the key analysis tool is here the average of the fictitious barrier widths $\xi^\nu_d$, which naturally follows the mean size of symmetry domains.
Indeed, larger domain size $N_d$ allows for larger $\xi^\nu_d$ corresponding to states which stochastically localize further from the domain center.
This in turn favors the formation of $lr$ pairs from boundary-perturbed combinations of $eo$ pairs, as described above.

\subsection{Statistical eigenstate spatial properties \label{sec:statAnalysis}}

\begin{figure}[t!] 
\center
\includegraphics[width=.99\columnwidth]{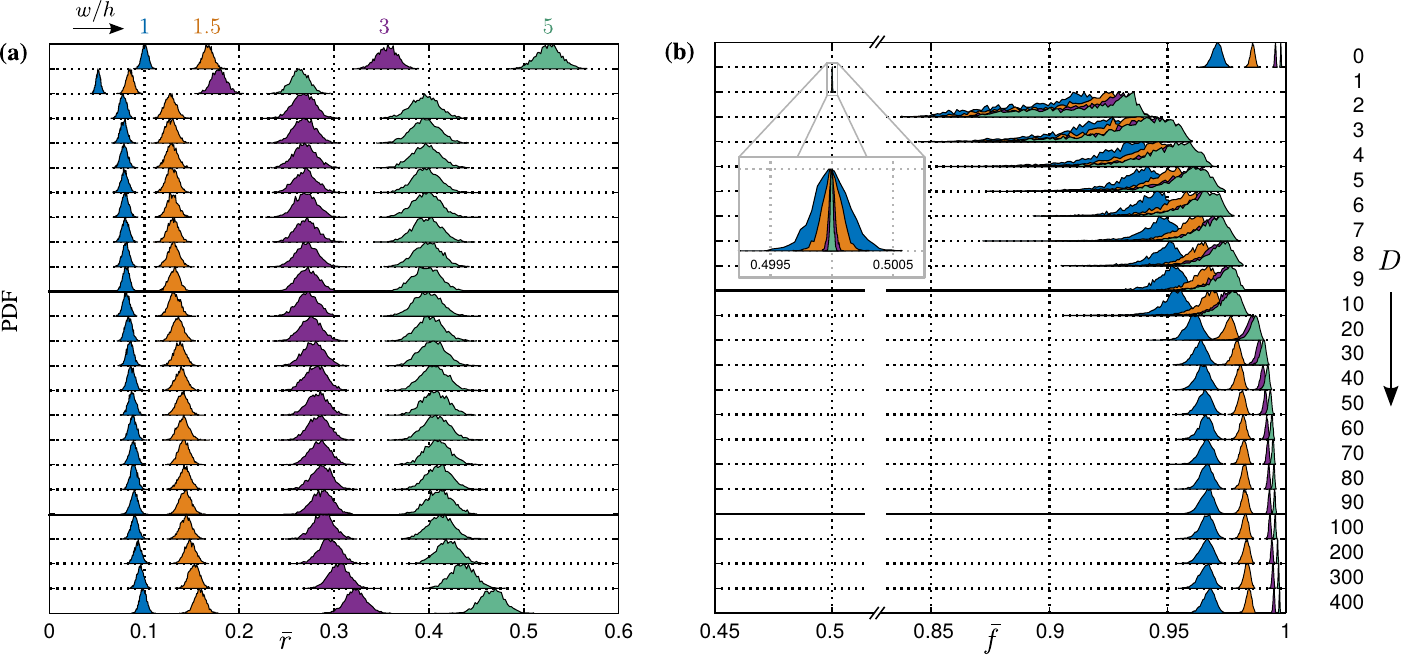}
\caption{
\textbf{(a)}: Probability density function (PDF) of mean inverse participation ratio (IPR) $\bar{r}$ over the $N = 500$ eigenstates of a locally reflection-symmetric disordered (LRD) chain, for varying disorder strength $w$ and number $D$ of (randomly sized) symmetry domains, using $3000$ random configurations.
\textbf{(b)}: Same as in (a), but for the mean cumulative Friedel sum (CFS) $\bar{f}$.
PDFs are normalized to maximum for each $(w,D)$ pair.
}
\label{fig:ipr_cfs}
\end{figure}

With the above insight into individual eigenstate profile characteristics, we now analyze the statistical behavior of eigenstate localization in LRD chains for varying disorder and setup symmetrization. 
To this end, we compute \cite{MpackGMP} the probability distribution function (PDF) of the mean IPR $\bar{r}$ and CFS $\bar{f}$ over the eigenstates of a given configuration, where $\bar{x} = \sum_{\nu=1}^N x_\nu / N$ with $x = r, f$.
The result is shown in \cref{fig:ipr_cfs} for different disorder strengths $w$ and number of symmetry domains $D$.
As we see, for each $(w,D)$-combination the IPR and CFS distributions have well-defined single maxima.
Note that $D = 0$ represents a random chain without any symmetrization and $D = 1$ a globally symmetric disordered chain, while the maximal value $D = N$ (not shown) is equivalent to $D = 0$.

As expected, we see in \cref{fig:ipr_cfs}\,(a) that the mean IPR is peaked at higher $\bar{r}$---that is, eigenstates are more localized---at stronger disorder, for any number of symmetry domains.
Indeed, the Anderson localization mechanism will govern the spatial decay on the single-site length scale within the region where a state is concentrated, independently of the presence of symmetries on larger scales.
Larger $w$ then leads to faster decay and larger $\bar{r}$ on average.
At the same time, the fluctuations around the peak $\bar{r}$ value (width of each PDF hump) increase with $w$, since individual $r$-values---being quadratic in $\rho_n$---are more sensitive to detailed differences between spatial configurations for more localized (larger $\rho_n$-values) states.

On top of those short-range statistical characteristics, \cref{fig:ipr_cfs}\,(a) shows a systematic impact of the long-range spatial correlations of the chain on the IPR distributions induced by local symmetry (we comment explicitly on the global symmetry case $D = 1$ later):
For any disorder strength $w$, the PDFs shift to lower $\bar{r}$ when adding local symmetry (from $D = 0$ to any $D \neq 0$), and rise to higher $\bar{r}$ as $D$ increases, approaching the random limit again at $D = N$.
Let us analyze the PDF evolution with varying $D$ more specifically as a result of the local symmetry-induced state profiles described in \cref{sec:symmetrization}.
Since increasing $D$ yields smaller domains on average, it thereby also leads to smaller average widths $\xi^\nu_d$ of the fictitious double-well barriers (see \cref{fig:selectedStates}\,(a)) induced by the disorder.
This in turn favors the occurrence of $eo$ state pairs (with relatively small IPR) compared to $lr$ pairs (with relatively large IPR), as concluded in \cref{sec:symmetrization}.
At the same time, however, the occurrence of domain-interface-localized asymmetric states (of type (iv) in \cref{sec:symmetrization}), of relatively large IPR, increases for larger number ($D-1$) of domain interfaces.
Also, $eo$ pairs become less supported again for smaller domains where the states do not have enough available space to localize away from the parity-breaking domain boundaries.
Together, these effects lead to a gradual increase of the mean IPRs for larger $D$ towards the limit $D \to N$.

More striking aspects related to the spatial state profiles for different $D$ and $w$ are captured by the CFS distribution shown in \cref{fig:ipr_cfs}\,(b).
The short-range localization behavior of the mean CFS $\bar{f}$ is similar to that of the IPR discussed above:
For given $D$, the peak $\bar{f}$ values increase with disorder strength $w$, indicating stronger localization.
Interestingly, however, the fluctuations around the peaks show a behavior opposite to the IPRs at $D \gtrsim 10$, now being more enhanced for weaker disorder.
This indicates that the average degree of spatial fragmentation (as measured by $\bar{f}$) is more homogeneous for large $w$ with stronger short-range localization, while smaller $w$ favors fragmentation variability for the more smeared out states.
Also, the $\bar{f}$-fluctuations increase for smaller number of domains (towards $D = 2$), since there the range of variability in the domain sizes $N_d$ increases.

The evolution of the CFSs with varying domain number $D$ generally follows the same scheme as the IPRs, with an initial shift to smaller $\bar{f}$ with the onset of local symmetry and a gradual recovering of the random limit as $D \to N$.
Complementing the IPRs as an eigenstate fragmentation measure, the CFSs reveal the role of $eo$ state pairs in a more resolved manner with increasing $D$.
Specifically, smaller domains (smaller fictitious intra-domain barrier widths $\xi^\nu_d$) on average favor the occurrence of $eo$ over $lr$ pairs, and thus the CFSs systematically increase with the decreasing fragmentation of $eo$ states.
At the same time, the fluctuations around the CFS peaks decrease with $D$ since the smaller domains allow for smaller variation in the extent of symmetrized states.

For a more complete picture of the above features, statistical distributions of the number of locally approximate even/odd and left-/right-localized states (including $eo$- and $lr$-pairs) are given in \cref{app:eigenstateSymmetryStat} for varying $w$ and $D$.
They demonstrate a gradual diminishing of domain-localized states with increasing $D$, though at a slower rate for locally even/odd than for left-/right-localized states.

An extreme manifestation of symmetry-induced localization in the IPR and CFS distributions naturally occurs for global symmetry ($D = 1$), where all eigenstates are even or odd.
With a fixed domain size $N_1 = N$, the majority of eigenstates generally consists of $eo$ pairs, inducing a dramatic jump of the PDFs to lower $\bar{r}$ and $\bar{f}$.
In particular, $\bar{f}$ is sharply peaked at $1/2$ for all disorder strengths, indicating an average $eo$-state density peak spacing $\xi$ of $N/2$ (recall the behavior of $f$ from \cref{sec:IPRvsCFS}), with fluctuations decreasing with $w$.
This feature is analyzed in detail in \cref{app:noVersusGlobalSymmetry}, where also the dependence on the chain length is investigated.

Finally, it should be noted that, whereas the local symmetries have substantial impact on the chain eigenstates for different $D$, the corresponding eigenenergy spectra share the overall trend of uncorrelated random chains.
More specifically, for sufficiently strong disorder the eigenenergies $\epsilon_\nu$ tend to increase linearly with $\nu$, as seen in \cref{fig:setupExample} on the scale of the plot.
The difference from a non-symmetrized chain lies in the occurrence of multiple quasidegeneracies on smaller scale (discernible when stretching the $\epsilon$-axis) corresponding to $eo$ or $lr$ state pairs, as described above.
A more detailed analysis of the spectrum of this exemplary LRD chain (with $D = 10$ domains) is given in \cref{app:spectralStat}, together with the ensemble average thereof.

\section{Transfer efficiency in LRD chains with adjacent and overlapping symmetry \label{sec:dynamics}}

Having investigated and explained the static eigenstate properties of LRD chains, let us now explore the impact of local symmetry on the dynamics of evolving wave-packets.
Since any wave-packet will evolve according to its projection coefficients on the chain's eigenstates, the question of interest here will be whether the occurrence of approximate local parity eigenstates may systematically affect the dynamics.

\begin{figure}[t!] 
\center
\includegraphics[width=.7\columnwidth]{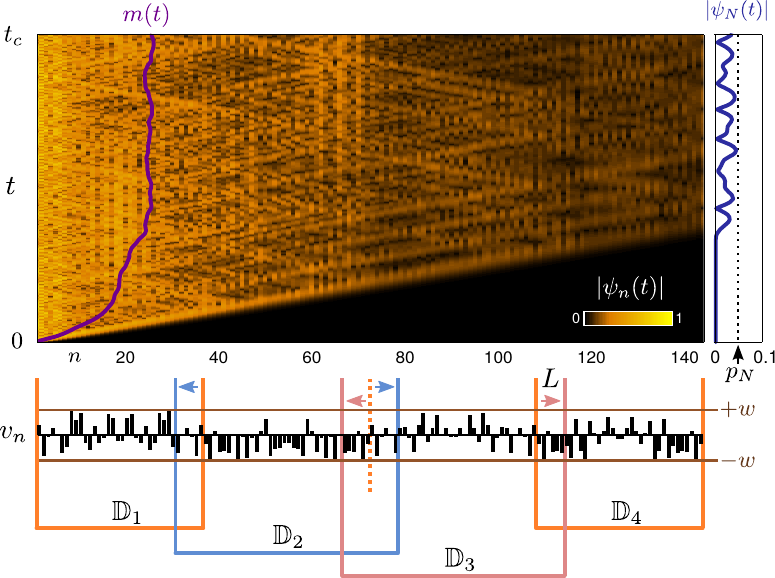}
\caption{
Wave-packet evolution $|\psi_n(t)|$ up to $t_c = 200/h$ with superimposed mean displacement $m(t)$ (top) upon unit excitation at site $n = 1$ of an LRD chain of disorder strength $w = h$ with $D = 4$ overlapping symmetry domains (bottom) where $\DD_2$ and $\DD_3$ have been extended symmetrically by $L$ sites.
The transfer efficiency $p_N$ from first to last site is indicated (top right).
}
\label{fig:evolution}
\end{figure}

In the study of correlation-induced effects on wave diffusion during evolution, a widely used measure is the root-mean-square (standard) deviation, defined as \cite{Sales2012_PELSaN_45_97_NumericalStudyOneelectronDynamics} 
\begin{equation} \label{eq:rmsd}
 m(t) = \sqrt{\sum_n (n - n_0)^2 |\psi_n(t)|^2}
\end{equation}
for a discrete chain, where the wave-packet $\ket{\psi(t)}$ evolves here according to the Schr\"odinger equation (with $\hbar \equiv 1$)
\begin{equation} \label{eq:evolution}
 i \partial_t \ket{\psi(t)} = H \ket{\psi(t)}
\end{equation}
under an initial unit excitation at the single site $n_0$, $\ket{\psi(t=0)} = \ket{n_0}$.
Such an evolution is shown in \cref{fig:evolution} for initial excitation at the left end $n_0 = 1$ of an LRD chain.
For simplicity, we have chosen a relatively short chain with a small number $D = 4$ of symmetry domains.
In this example, however, we also introduce a spatial \textit{overlap} between consecutive domains which, as we will see below, may drastically affect the diffusion.
Specifically, in \cref{fig:evolution} the overlap is created by extending each domain except for the first ($\DD_1$) and last ($\DD_D$) by a fixed number of sites $L$, keeping the domain centers fixed.

The mean displacement $m(t)$ for the configuration in \cref{fig:evolution} increases with time, as expected, until $\psi_n(t)$ has spread enough to reach the right end where it is back-reflected.\footnote{Since we here explicitly investigate transfer in \textit{finite} chains, we let the wave-packet reach the end of the chain, leading to a saturation of $m(t)$. This is in contrast to studies on approximants of infinite chains, where back-reflection of the wave is avoided and $m(t)$ increases as $\sim \sqrt{t}$ on average for uncorrelated disorder.}
From this point on $m(t)$ simply fluctuates around a constant mean value.
This is, nevertheless, the generic behavior also for uniformly random chains without local symmetry correlations, with the saturation mean value for $m(t)$ decreasing with disorder strength.
For other types of (short- or long-range) correlation, the effect on the rate of increase of $m(t)$ is usually studied before reflection at the end of the chain sets in \cite{Dunlap1990_PRL_65_88_AbsenceLocalizationRandomdimerModel}.
Local-symmetry-induced correlations, however, do not affect the overall displacement behavior on ensemble average:
In analogy to the IPR used in \cref{sec:eigenstateLocalization}, the mean displacement does not resolve details of the time-dependent spatial profile of $|\psi_n(t)|$ (like, e.\,g., the faint but visible slight enhancement of $|\psi_n(t)|$ close to the overlap between domains $\DD_2$ and $\DD_3$ seen in the color-plot of \cref{fig:evolution}).

To probe possible symmetry-induced dynamical effects in the LRD chain in a site-resolved manner, we will utilize the so-called ``transfer efficiency'' \cite{Walschaers2015_PRE_91_42137_StatisticalTheoryDesignedQuantum,Zech2014_NJP_16_55002_CentrosymmetryEnhancesQuantumTransport} of the initial excitation to site $n$, defined here as the maximum amplitude at $n$ over a fixed reference time $t_c$,
\begin{equation} \label{eq:transfEff}
 p_n = \max_{t \in [0,t_c]}{| \langle n | e^{-iHt} | n_0 \rangle |} = \max_{t \in [0,t_c]}|\psi_n(t)|~ \in [0,1],
\end{equation}
where we set the input site to $n_0 = 1$.
To visualize an example, $p_{n=N}$ is indicated in \cref{fig:evolution}\,(right panel) for that setup. 
The transfer efficiency has been used to demonstrate that \textit{global} symmetry in discrete disordered networks may generally lead to an enhanced signal transmission between diametrically located input and output end-sites \cite{Zech2014_NJP_16_55002_CentrosymmetryEnhancesQuantumTransport}.
This effect relies on the commutation of the Hamiltonian with the global reflection operation, and was shown to be subject to further conditions and optimizations when promoted for efficient quantum transport \cite{Walschaers2015_PRE_91_42137_StatisticalTheoryDesignedQuantum}.

\subsection{Transfer enhancement by local symmetry \label{sec:transfEff}}

\begin{figure}[t!] 
\center
\includegraphics[width=.9\columnwidth]{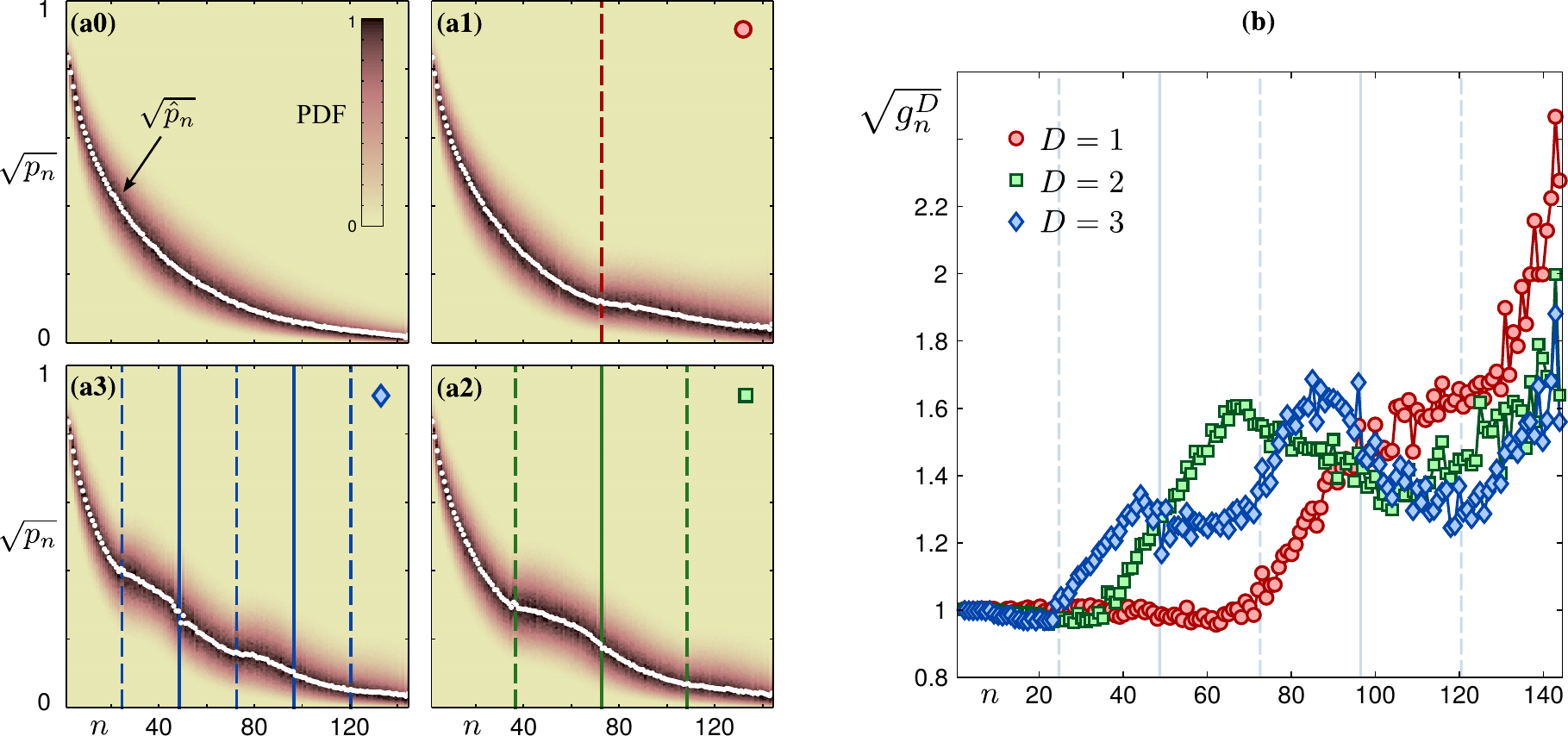}
\caption{
PDFs of scaled transfer efficiency $p_n$ (normalized to maximum for each $n$) within cutoff time $t_c = 200/h$ ($1000$ time-steps), for an ensemble of $30\,000$ LRD chains ($N = 144$, $w = 1.2\,h$) with 
\textbf{(a0)} no symmetry ($D = 0$), \textbf{(a1)} global symmetry ($D = 1$), \textbf{(a2)} $D = 2$ and \textbf{(a3)} $D = 3$ adjacent symmetry domains, with PDF peak positions $\hat{p}_n$ (white dots) estimated using local regression.
Solid (dashed) vertical lines indicate domain interfaces (centers).
\textbf{(b)} Transfer efficiency enhancement factor $g_n^D = \hat{p}_n^D / \hat{p}_n^0$ (scaled) for $D = 1,2,3$.
}
\label{fig:transfEff}
\end{figure}

What we aim to investigate here is whether a statistical enhancement of signal transfer (compared to uncorrelated disorder) can be manifest if more than one \textit{local} symmetries are present in the finite chain, each of which now does not commute with $H$.
To this end, we first consider the case of adjacent, that is, non-overlapping symmetry domains.
We compute the PDFs of the site-resolved transfer efficiency $p_n$ over an ensemble of disordered configurations of an $N=144$-site chain with $D = 1,2,3$ symmetry domains, and compare it to the non-symmetric, uniformly random case ($D = 0$).
The results are shown in \cref{fig:transfEff}\,(a0)--(a3), where $\sqrt{p_n}$ is plotted to increase detail visibility.
We use here a disorder strength $w = 1.2\,h$ and evolution time $t_c = 200/h$ in \cref{eq:transfEff} such that the wave-packet has explored the whole chain.
As we see, in all cases the PDF for any given site $n$ is rather peaked (with peaks becoming narrower towards the chain ends), and the peak $p_n$-values fall monotonously with $n$, as can be anticipated for a disordered chain.
Notably, one can clearly distinguish a relatively small but statistically systematic enhancement of $p_n$ when local symmetry is imposed, approximately in the right halves of the symmetry domains; 
see local humps of PDF peaks along $n$ in \cref{fig:transfEff}\,(a1),(a2),(a3).

For a clearer comparison with the non-symmetric case (a0), \cref{fig:transfEff}\,(b) shows (scaled) the enhancement quotient 
\begin{equation} \label{eq:enhancement}
 g_n^D = \frac{\hat{p}_n^D}{\hat{p}_n^0}
\end{equation}
of the peak transfer efficiencies $\hat{p}_n^D$ (corresponding to $D = 1,2,3$ domains) to that of the uncorrelated random chain, $\hat{p}_n^0$.
The PDF peak values (white dots in \cref{fig:transfEff}\,(a0)--a3)) have been estimated as the maxima of smoothed versions of the PDFs using local regression \cite{Marsh2018_LOESS}.
The $g_n^D$ show a degree of fluctuation increasing along $n$, which stems from the strong interference-induced $p_n$-fluctuations among individual configurations.
They clearly demonstrate, though, an \textit{enhancement in transfer efficiency when local symmetry is added} ($g_n^D > 1$), to chain parts dependent on the symmetry domains.

For $D = 1$ (global symmetry), the enhancement practically starts when crossing the symmetry center, and is then steadily increased in the right chain half.
In similarity to the network case of Ref.\,\cite{Zech2014_NJP_16_55002_CentrosymmetryEnhancesQuantumTransport}, this is a consequence of the definite parity of the eigenstates under global reflection:
When those parity eigenstates have a finite projection onto the initial input state (on the left chain half), they will enable tunneling to their mirror-related part (on the right half).
The total transfer is determined by the combination of such effective ``double-well'' tunnelings \cite{Diez2008_PRB_78_35118_SymmetryinducedTunnelingOnedimensionalDisordered}.
For $D = 2$, the two local reflection operations $\P_{\DD_{1,2}}$ do not commute with $H$, but still there is a multitude of \textit{approximate} local parity eigenstates (see \cref{sec:symmetrization}) which can assist in tunneling between the two halves of each domain.
Indeed, we observe a drastically enhanced transfer to the right half ($n \in [36,72]$) of $\DD_1$ which stops roughly at the interface to $\DD_2$.
Then $g_n$ slightly drops, and increases again in the right half ($n \in [108,144]$) of $\DD_2$.
The scheme of increased enhancement in right domain halves is similar for $D = 3$.
Note that the $g_n$-curves are scaled by the domain size:
Their slopes are approximately the same in corresponding domain parts, with the slopes overall decreasing towards the right chain end.

The main difference to the global symmetry case is a generally significant portion of $lr$ pair states as well as asymmetric domain-interface-localized states, depending on the fictitious intra-domain barriers (see \cref{sec:symmetrization}).
These states do not contribute to the intra-domain tunneling and therefore lower the transfer enhancement compared to the global symmetry case.
For the weak disorder chosen in \cref{fig:transfEff}, the occurrence of such states is overall reduced, but at the same time \textit{extended} asymmetric states are favored.
Those may generally contribute to transfer, though also in the non-symmetric chain, and are therefore not expected to increase the $g_n^D$.

\subsection{Overlap-induced transfer enhancement \label{sec:transfEffOverlap}}

Finally, an intriguing variation on the above LRD setups is to introduce spatial overlap between the domains $\DD_d$ (as in the explicit example of \cref{fig:evolution}).
We remark that such \textit{domain overlap is a unique characteristic accessible with local symmetry} as opposed to global symmetry \cite{Morfonios2017_AoP_385_623_NonlocalDiscreteContinuityInvariant}.
The key feature here is that symmetry-adapted LRD chain eigenstates (that is, $eo$ and $lr$ pairs as well as single approximate local parity eigenstates; see \cref{sec:symmetrization}) of one domain may have substantial spatial overlap with those of a consecutive domain within the overlap region.
Since an evolving wave-packet generally has contributions from all available eigenstates, it may be transferred across domains via this spatial overlap of different eigenstates.

\begin{figure}[t!] 
\center
\includegraphics[width=.9\columnwidth]{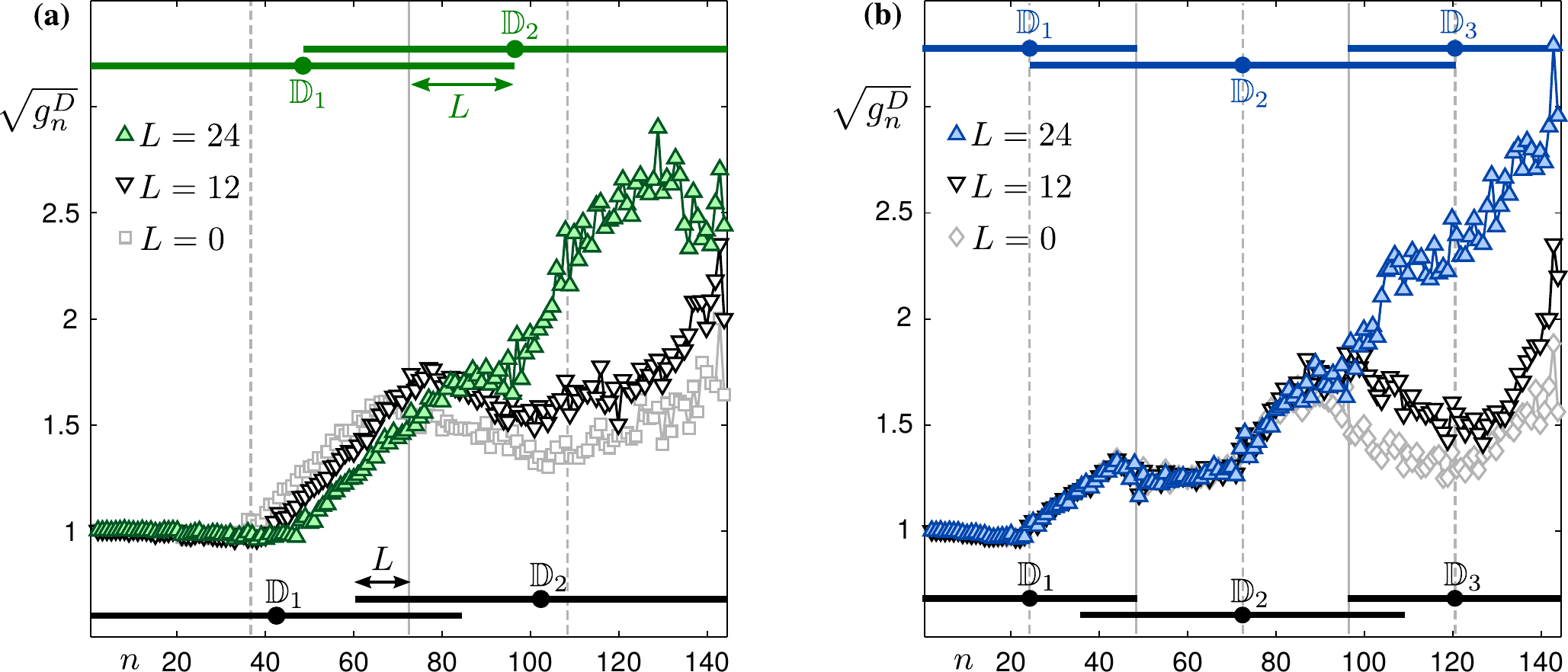}
\caption{
Transfer efficiency enhancement factor $g_n^D$ (scaled) like in \cref{fig:transfEff}\,(b) but for finite overlap between symmetry domains created by 
\textbf{(a)} extending each of $D = 2$ domains by $L$ sites across the chain center and \textbf{(b)} extending the middle one of $D = 3$ domains symmetrically by $L$ to the left and right.
Horizontal bars and dots indicate domains and their centers for $L = 12$ (bottom) and $L = 24$ (top).
Solid (dashed) vertical lines indicate domain interfaces (centers) for $L = 0$.
Gray curves for $L = 0$ in (a) and (b) are the same as those in \cref{fig:transfEff}\,(b) with same markers (green and blue, respectively).
}
\label{fig:transfEffRatioOverlap}
\end{figure}

Two such scenarios are realized in \cref{fig:transfEffRatioOverlap} for (a) $D = 2$ and (b) $D = 3$ symmetry domains, in two different ways (see horizontal bars indicating domains):
In \cref{fig:transfEffRatioOverlap}\,(a) both domains are extended by $L$ sites across the middle of the chain (with equal domain sizes $N_1 = N_2 = N/2$ for $L = 0$), with their centers shifted by $L/2$, while in \cref{fig:transfEffRatioOverlap}\,(b) only the middle one of three equally sized domains is extended symmetrically by $L$, such that the domain centers remain fixed.
In both cases, we observe a clear enhancement of transfer to the right half of the whole chain in the presence of domain overlap compared to adjacent domains ($L = 0$, gray curves). 

Note that for $D = 2$, the overlap leads to a so-called \textit{gapped translation symmetry} \cite{Kalozoumis2014_PRL_113_50403_InvariantsBrokenDiscreteSymmetries,Morfonios2017_AoP_385_623_NonlocalDiscreteContinuityInvariant}: The chain along the first $2L$ sites is repeated in (i.\,e., finitely ``translated'' to) the last $2L$ sites, but not in the region between which constitutes a symmetry ``gap''; note though, that the mirror image of the translated part appears within the domain overlap.
For $D = 3$, the overlap yields a \textit{gapped reflection symmetry} \cite{Kalozoumis2014_PRL_113_50403_InvariantsBrokenDiscreteSymmetries,Morfonios2017_AoP_385_623_NonlocalDiscreteContinuityInvariant}:
the chain is reflection-symmetric about its center, with the exception of the $N_d-2L$ sites around the centers of $\DD_{1}$ and $\DD_3$ forming a (locally symmetric) gap.
As it appears, those long-range correlations induced via overlap-induced gapped symmetries may play a substantial role in enhancing signal transfer through LRD systems.
An interesting prospect would be to explore their impact for larger number of overlapping domains (of same or different sizes) featuring multiple symmetry gaps.

In the special case of $L = N/6$, for both ($D = 2,3$) of the two considered LRD chain setups in \cref{fig:transfEffRatioOverlap} there is a dramatic enhancement of the ensemble-average transfer efficiency, with maximal $g_n$-factors reaching $g_n \approx 9$ (almost double the average maximal $g_n$ for the globally symmetric setup in \cref{fig:transfEff}\,(b)).
Now, the $D = 2$ chain consists of a single part (first $2L$ sites) which is successively reflected two times at its right end, while the $D = 3$ chain becomes globally symmetric but additionally composed of two different symmetric units of size $2L$ (one in the middle and one repeated at the two ends).
In particular, the latter case indicates that the possible transfer efficiency enhancement by global symmetry \cite{Zech2014_NJP_16_55002_CentrosymmetryEnhancesQuantumTransport} may be even further \textit{increased drastically if local symmetry is present simultaneously at smaller scales} within a composite system.

\section{Conclusions \label{sec:conclusions}}

We have investigated the localization and signal transfer properties of finite, locally reflection-symmetric disordered (LRD) tight-binding chains, treating local symmetry as a spatial correlation of variable range.
To reveal the localization behavior, we used the ensemble distributions of the inverse participation ratio (IPR) and a recently proposed measure of confinement here coined ``cumulative Friedel sum'' (CFS).
It was shown that the spatial participation and fragmentation of eigenstates increases in the presence of local symmetries, and decreases towards the limit of uncorrelated disorder for increasing number of randomly sized symmetry domains, with statistical distributions depending on the disorder strength.
The localization behavior is induced by the disordered symmetry domains acting as fictitious double wells in which eigenstates acquire approximate local parity.
This type of symmetrized localization is explained within a local resonant scattering picture combined with the recent theory of effective confinement potentials.
Further, the dynamics of a wave-packet upon excitation of the leftmost site in LRD chains was investigated in terms of the site-resolved transfer efficiency.
Here, a systematic enhancement of transfer to the right halves of one, two, or three symmetry domains was shown to take place compared to the non-symmetric random chain.
This enhancement can be drastically increased in the presence of overlap between symmetry domains; especially in the case of repeated extended constituents in the chain, or in the simultaneous presence of global and local reflection symmetry.
In particular, the possibility to amplify signal transfer by the coexistence of global and local symmetry in composite systems is thus demonstrated. 

We stress that the aim of the present work is to investigate the generic impact of the presence of local symmetry on localization and state transfer efficiency in a minimalistic setting.
Disordered 1d chains with uniformly random potential were thus chosen as a platform to isolate the effect solely of the imposed symmetry---that is, without the influence of other structural characteristics or assumptions.
We also underline that our results concern the properties of \textit{finite} LRD chains, and are not to be seen as a study of correlations in approximants of the $N \to \infty$ limit; chain sizes were simply chosen large enough to perform statistics and vary the number or size of symmetry domains.

Certainly, many alternative routes could be employed to optimize the parameters of LRD setups for efficiency or to probe the effect of local symmetries with improved symmetry-adapted measures.
The insight provided here may then also be leveraged to \textit{design} devices with (overlapping) local symmetries, in order to achieve controllable localization or signal transfer at desired locations.
As an example, we mention the perspective to combine the concept of overlapping local symmetries with the phenomenon of so-called ``necklace states'' \cite{Bertolotti2005_PRL_94_113903_OpticalNecklaceStatesAnderson,Sgrignuoli2015_AP_2_1636_NecklaceStateHallmarkDisordered} of spatially and spectrally overlapping resonances to achieve, e.\,g., simultaneous spatial confinement and transmission control.
Alternatively, local symmetries may be applied to special types of structured disorder enabling engineered wave transport \cite{Rodriguez2012_PRB_86_85119_ControlledEngineeringExtendedStates}.
With the present work we take a step in the direction of understanding and manipulating the effect of coexisting local symmetries in a medium on its wave response, a concept which can be modeled in more complex systems and in higher dimensions.

\section{Acknowledgements}

P.\,S. acknowledges financial support by the Deutsche Forschungsgemeinschaft under grant DFG Schm 885/29-1. 
M.\,R. is thankful to the `Stiftung der deutschen Wirtschaft' for financial support in the framework of a scholarship.

\appendix

\section{Local resonant scattering picture: localization and symmetrization from transparency \label{app:localScattering}}

At the heart of the present study of LRD chains lies the tendency of eigenstates to localize into and symmetrize within symmetric domains, as described in \cref{sec:symmetrization}---with asymmetric (but mirror-related) $lr$ pair states resulting from combined quasidegenerate $eo$ pair states perturbed by domain boundaries.
We now give an intuitive argument for the symmetrized eigenstate localization in LRD chains, based on the combination of a recent unifying theory of wave localization \cite{Filoche2012_P_109_14761_UniversalMechanismAndersonWeak} with a scattering picture of perfectly transmitting \cite{Kalozoumis2013_PRA_88_33857_LocalSymmetriesPerfectTransmission,Morfonios2017_AoP_385_623_NonlocalDiscreteContinuityInvariant} local resonant states.
We split the argument by answering three questions, as follows.

\begin{figure}[!ht] 
\center
\includegraphics[width=.6\columnwidth]{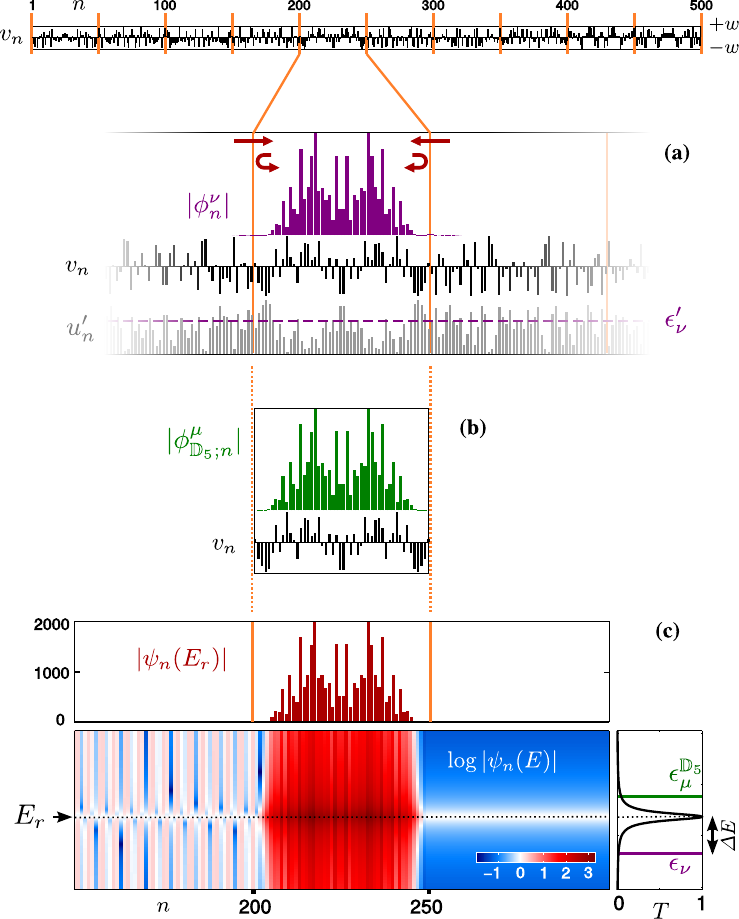}
\caption{
\textbf{(a)}: Focus on eigenstate $\phi^{\nu=360}_n$ localized within domain $\DD_5$ of an LRD chain (complete potential $v_n$ in \textit{top inset}, with domain interfaces indicated by vertical orange lines) with disorder strength $w = 3h$ and $D = 10$ domains of fixed size $N_d = N / D = 50$, together with dual effective confining potential $u_n'$ (gray) and dual eigenenergy $\epsilon_{\nu=360}'$ (dashed horizontal line); see \cref{eq:landscape} and text below.
\textbf{(b)}: Eigenstate $\phi^{\mu=36}_{\DD_5;n}$ of isolated domain $\DD_5$ (truncated chain potential shown in bottom).
\textbf{(c)}: Perfectly transmitting ($T(E_r) = 1$) scattering state $\psi_n(E_r)$ (\textit{top}) for plane wave incident on isolated domain $\DD_5$ at resonant energy $E_r = \epsilon_{360} + \varDelta E$ with $\varDelta E \approx 3.6\times 10^{-7}\,h$, and color-plot of scattering state map $\psi_n(E)$ on logarithmic scale (\textit{bottom}) as well as transmission $T(E)$ (\textit{bottom right}) in the vicinity of $E_r$, with eigenenergies $\epsilon_{\nu=360} = 1.265367\,h$ (of $\phi^{\nu=360}_n$) and $\epsilon_{\mu=36}^{\DD_5} \approx \epsilon_{360} + 1.55\, \varDelta E$ (of $\phi^{\mu=36}_{\DD_5;n}$) indicated by horizontal lines.
The transparency of $\DD_5$ at $E_r$ is indicated by straight arrows in (a), while round arrows indicate back-reflection from the neighboring domains, effectively leading to localization of $\phi^{\nu}_n$ into $\DD_5$ in a local scattering picture (see \cref{sec:transparency}).
State moduli are plotted in arbitrary units in (a,b) and normalized to unit incoming flux in (c).
}
\label{fig:localScattering}
\end{figure}

\subsection{Where in a disordered chain can an eigenstate localize?}

To begin with, computing the eigenvectors of \cref{eq:evp} for a generic disordered medium raises the question:
What determines the positions and ranges of localization corresponding to given eigenvalues?
The answer is provided in the fairly recently developed framework of ``effective confining potentials'' \cite{Arnold2016_PRL_116_56602_EffectiveConfiningPotentialQuantum} and ``localization landscapes'' \cite{Filoche2012_P_109_14761_UniversalMechanismAndersonWeak}, formulated also for discrete models \cite{Lyra2015_E_109_47001_DualLandscapesAndersonLocalization}.
We now briefly outline this framework, and provide an example for an LRD chain in \cref{fig:localScattering}\,(a) (see below).

For our discrete chains, the so-called \textit{effective confining potential} $u_n$ is defined as the inverse of the ``landscape function'' $\tau_n$, in turn given as the site amplitudes of the response $\ket{\tau}$ (solving $H_s \ket{\tau} = \ket{e}$) of the system to a spatially uniform excitation (source term) $\ket{e}$ \cite{Arnold2016_PRL_116_56602_EffectiveConfiningPotentialQuantum,Lyra2015_E_109_47001_DualLandscapesAndersonLocalization}:
\begin{equation} \label{eq:landscape}
 u_n = \frac{1}{\tau_n} = \frac{1}{\braket{n|\tau}}, \quad \ket{\tau} = H_s^{-1} \ket{e}
\end{equation}
with $e_n = 1$ $\forall n$, where $H_s = H + V_s$ with a constant offset diagonal $V_s$ added such that $u_n > 0$.
As shown in Ref.\,\cite{Arnold2016_PRL_116_56602_EffectiveConfiningPotentialQuantum}, an eigenstate $\ket{\phi^\nu}$ of $H$ with eigenenergy $\epsilon_\nu$ decays exponentially within regions $n$ where $\epsilon_\nu < u_n$, and can thus have substantial amplitude only in the remaining regions---that is, within local minima of $u_n$ below the threshold $\epsilon_\nu$.
In other words, $u_n$ defines the locations to which $\phi^\nu_n$ can be \textit{spatially confined according to its energy}, namely between ``effective barriers'' where $\epsilon_\nu < u_n$.
At larger $\epsilon_\nu$ the eigenstate will be less localized, since such barriers between local $u_n$ minima are exceeded and larger regions are available.

For discrete (tight-binding) models, eigenstates localize again for higher energies (like, e.\,g., state $\ket{\phi^{489}}$ in \cref{fig:selectedStates}\,(a)), though now confined by a so-called ``dual'' effective potential $u_n'$.
It is obtained by using $H_s' = V_s' - H$ (with eigenenergies $\epsilon_\nu'$) instead of $H_s$ in \cref{eq:landscape} where, again, the constant offset $V_s'$ is added to have $u_n' > 0$.
This is shown in \cref{fig:localScattering}\,(a) for a relatively high-energy eigenstate localizing into domain $\DD_5$ of a LRD chain---here an example with $N = 10$ equally sized domains.
The state is indeed confined between two thick effective barriers of the corresponding $u_n'$ (where $\epsilon_\nu' < u_n'$) close to the borders of the domain.
Smaller barriers lead to amplitude minima in the domain interior.
Note here that $u_n$ (and $u_n'$) follows the local symmetry of the original potential $v_n$.

\subsection{Which possible location does an eigenstate choose to localize in?}

Clearly, for a given level $\epsilon_\nu$, there are \textit{multiple} local $u_n$ (or $u_n'$) minima which could host the corresponding eigenstate; 
see e.\,g. the $\epsilon_\nu' < u_n'$ regions surrounding $\DD_5$ in \cref{fig:localScattering}\,(a).
As shown in Ref.\,\cite{Filoche2012_P_109_14761_UniversalMechanismAndersonWeak} (and in a recent extension to discrete models \cite{Rontgen2018_AC___LocalSymmetryTheoryResonator}), the position where the eigenstate will actually localize is then determined by the minimal spectral distance 
\begin{equation} \label{eq:spectralDistance}
 \delta\epsilon = \min_{\CC,\mu}|\epsilon_\nu - \epsilon^\CC_\mu|
\end{equation}
of the eigenenergy $\epsilon_\nu$ to the eigenspectra $\{ \epsilon^\CC_\mu \}$ of sub-Hamiltonians $H_\CC$ of (possible) domains of localization $\CC$:
The smaller $\delta\epsilon$ is for a given region $\CC$, the larger is the allowed norm of $\ket{\phi^\nu}$ (eigenvector of $H$) within $\CC$ for given boundary data (in the present discrete case, values of $\phi^\nu_n$ at the sites adjacent to $\CC$ \cite{Rontgen2018_AC___LocalSymmetryTheoryResonator}).
Particularly, in the limiting case of zero boundary data, $\phi^\nu_{n \in \CC} \neq 0$ only if $\epsilon_\nu = \epsilon^\CC_\mu$ for some $H_\CC$-eigenstate $\ket{\phi^\mu_\CC}$ under Dirichlet boundary conditions \cite{Filoche2012_P_109_14761_UniversalMechanismAndersonWeak}.
This essentially means that $\ket{\phi^\nu}$ will confine into the localization domain $\CC$ supporting a local $H_\CC$-eigenstate which best matches $\ket{\phi^\nu}$ in eigenenergy (i.\,e., with smallest $\delta\epsilon$).
It will then also match it in spatial profile, that is, with (approximately) locally symmetric $|\phi^\nu_n|$ for symmetric $v_{n \in \CC}$.
An example of this is given in \cref{fig:localScattering}\,(b), showing the local eigenstate $\ket{\phi^{\mu=36}_\CC}$ (where we have chosen $\CC = \DD_5$) matching $\ket{\phi^{\nu=360}}$ above of the full system which localizes in $\DD_5$.

As a side note, if the system contains repeated subdomains (not occurring in the present random potentials) such that corresponding repeated confining domains $\CC$ occur, then also the localization of an eigenstate will be repeated in those $\CC$'s (since their $|\epsilon_\nu - \epsilon^\CC_\mu|$ values will be equally small) with factors depending on the detailed configuration at those domains' boundaries.
This intuitively explains, e.\,g., the repeated amplitude patterns occurring in eigenstates of deterministic aperiodic structures \cite{Rontgen2018_AC___LocalSymmetryTheoryResonator} with correspondingly repeating sub-Hamiltonians, which feature abundant local symmetries at different scales \cite{Morfonios2014_ND_78_71_LocalSymmetryDynamicsOnedimensional}.

\subsection{Why is the chosen region of localization in the LRD chain symmetric? \label{sec:transparency}}

Even within the above localization framework, however, the ultimate question of eigenstate symmetrization remains:
Why does it happen that, for sufficiently strong localization, the domains $\CC$ with smallest $|\epsilon_\nu - \epsilon^\CC_\mu|$, where the full eigenstates are confined, coincide with symmetry domains of the LRD chains?

To give an intuitive answer, let us view a subdomain $\DD$ as a local scatterer within a generic chain, and consider the scattering of a monochromatic wave of energy $E$ incident from the left on the isolated $\DD$ connected to perfect semi-infinite chains (or ``leads'').
The transmission function $T(E) \in [0,1]$, which is independent of the side of incidence of the wave, gives the portion of the (unit) wave amplitude that transmits through $\DD$ and leaves the scatterer on the right, while the reflected part is given by $R = 1 - T$.
$T(E)$ naturally shows variations depending on the internal structure of $\DD$ and may, in particular, feature resonant peaks corresponding to quasibound states of the scatterer---with resonant widths proportional to the couplings of such states to the leads.
Crucially, now, an energetically isolated resonance always has \textit{perfect transmission} $T(E) = 1$ at the resonance position $E = E_r$ if the scatterer $\DD$ is symmetric, with resonant state spatial profile being symmetric \cite{Kalozoumis2013_PRA_87_32113_LocalSymmetriesOnedimensionalQuantum,Kalozoumis2013_PRA_88_33857_LocalSymmetriesPerfectTransmission}.
This is shown in \cref{fig:localScattering}\,(c) for scattering off the isolated domain $\DD_5$, which features a scattering resonance extremely close to the eigenenergy $\epsilon_{360}$ of the localized eigenstate in \cref{fig:localScattering}\,(a), with practically identical amplitude profile $|\psi_n(E_r)|$ (note the relative factor between $|\psi_{n \in \DD_5}|$ within the scatterer and $|\psi_{n \notin \DD_5}|=1$ within the leads).
This domain will thus be transparent at $E_r \approx \epsilon_{360}$ when embedded into the considered LRD chain, where eigenstates can be viewed as forming upon multiple scattering (and interference) of waves off local scatterers (as done also originally in, e.\,g., Anderson's work \cite{Anderson1958_PR_109_1492_AbsenceDiffusionCertainRandom}).
In other words, waves impinging from the left and right onto $\DD_5$ are let inside without reflection, while being reflected back into $\DD_5$ by adjacent domains when reaching its border from inside, as indicated by arrows in \cref{fig:localScattering}\,(a).
Thus, there will be an accumulation of amplitude in $\DD_5$ forming the localized eigenstate at $\epsilon_{360} \approx E_r$, while this eigenstate is expelled from other localization regions due to larger $|\epsilon_{360} - \epsilon_\mu^\CC|$, as discussed above.

The deviation from the above picture, that is, deviations of LRD chain eigenstate energies and profiles from local (perfectly transmitting) scattering resonance energies and profiles, increases with the leakage of the eigenstates through the symmetry domain boundaries.
This naturally occurs for smaller disorder strength $w$ relative to given eigenenergies, where disorder-induced spatial decay is weaker and, equivalently, more maxima in the effective confining potential $u_n$ are exceeded by the eigenenergies.

Summarizing, eigenstate symmetrization into symmetry domains will occur for strong enough disorder (yielding short-range decay at the scale of the domain sizes), at eigenenergies matching perfect transmission resonance energies of the corresponding isolated domains.
The link to the fictitious double wells defined in \cref{app:fictitiousBarriers} can be viewed as follows.
The effective confining potential of \cref{eq:landscape} governs the \textit{details} of localization of an eigenstate.
In the case of its symmetrization into a domain, its double-peak profile is represented by the simple picture of a fictitious double-well with corresponding strength and width.

\section{Fictitious eigenstate-specific double wells \label{app:fictitiousBarriers}}  

To provide a simple analysis tool relating the locally symmetrized eigenstates (see \cref{app:localScattering}) to the LRD chain characteristics, we here introduce an effective mapping of such eigenstates to corresponding local double wells.

Any finite piece of the disordered medium can be seen to act \textit{effectively} as a homogeneous potential barrier, in the sense that both may lead to a spatially exponential decay of an eigenstate.
In the uncorrelated disordered chain, an eigenstate $\ket{\phi^\nu}$ will typically be localized with an exponential decay of modulus envelope $\chi^\nu_n$, that is, $|\phi^\nu_n| \leqslant \chi^\nu_n \propto e^{-|n - n_\nu|/\ell}$, in both directions outwards from its maximum position denoted $n_\nu$.
Here, $\ell = \ell(\epsilon_\nu;w) \equiv 1/\gamma$ is the ``localization length'', defined as the inverse of the so-called Lyapunov exponent $\gamma$, which generally depends on $\epsilon_\nu$ and $w$ \cite{Rodriguez2006_JPAMG_39_14303_OnedimensionalModelsDisorderedQuantum}.
On the other hand, for a homogeneous periodic chain with onsite potential $v_n = v$ and dispersion relation $E = 2h\cos k + v$ of $H$ in \cref{eq:hamiltonian}, there are solutions exponentially decaying as $e^{-\kappa n}$ at imaginary momenta $k$ (with $ik \equiv \kappa \in \RR$) for energies $E$ outside the band, $|E-v| > 2h$.
We thereby associate an exponentially localized state $\ket{\phi^\nu}$ in the uncorrelated disordered chain with a constant \textit{fictitious potential barrier} of strength 
\begin{equation} \label{eq:fictitiousPotential}
 \tilde{v}^\nu \equiv \epsilon_\nu - 2h\cosh \gamma(\epsilon_\nu;w),
\end{equation}
that is, supporting decaying states with the same exponent $\kappa = \gamma(\epsilon_\nu;w)$ at $E = \epsilon_\nu$.

Spatially, this fictitious barrier starts roughly at the sites adjacent to the single site, denoted $n_\nu$, where $|\phi^\nu_n|$ is maximal.
In the LRD chain, however, if an eigenstate is symmetrized in a domain $\DD_d$, as described above, it has a second (local) maximum at the symmetry-related position $\P_{\DD_d}(n_\nu)$; see e.\,g. $\phi^{\nu:\min f}$ in \cref{fig:setupExample}.
In this case, the fictitious barrier acquires a \textit{finite width}, which we simply take to be the number of sites
\begin{equation} \label{eq:fictitiousWidth}
 \xi^\nu_d = |n_\nu - \P_{\DD_d}(n_\nu)| - 1 = 2 |n_\nu - c_d| - 1
\end{equation}
between the positions of the two symmetry-related local maxima of state $\ket{\phi^\nu}$ localized in $\DD_d$.
This is visualized in the example of \cref{fig:selectedStates}\,(a), showing also $\tilde{v}$ (superscripts dropped) for the selected state with an estimated $\ell = 1/\gamma = 0.8$ in \cref{eq:fictitiousPotential}.
As a comparison, also the corresponding (fictitious) localized state denoted $\varphi_n$ is shown, here produced by choosing a potential at $n_\nu$ and $\P_{\DD_d}(n_\nu)$ (with $\tilde{v}$ along the remaining chain) such that this state's energy matches $\epsilon_\nu$ \footnote{Recall that, in contrast to a continuous potential, the barrier strength $\tilde{v}^\nu$ need not exceed the eigenenergy for the eigenstate to decay exponentially: The eigenenergy here merely needs to lie outside the allowed band of real Bloch momenta for a homogeneous chain with potential $v_n = \tilde{v}^\nu$, that is, to fulfill the condition $|E-v| > 2h$. In particular, $\tilde{v}^\nu$ may also well be negative (like in \cref{fig:selectedStates}\,(a)), and therefore we call it ``strength'' instead of ``height''.}.

In the above situation, the interior of the symmetry domain $\DD_d$ effectively plays the role of a \textit{symmetric double well} (like the globally symmetric setup of Ref.\,\cite{Diez2008_PRB_78_35118_SymmetryinducedTunnelingOnedimensionalDisordered}, but here coupled to adjacent sites), with a constant tunneling barrier of width $\xi^\nu_d$ and strength $\tilde{v}^\nu$.
The outer ``walls'' of this fictitious double well are represented by the constant potential $\tilde{v}^\nu$ on the left and right of $n_\nu$ and $\P_{\DD_d}(n_\nu)$, respectively (see \cref{fig:selectedStates}\,(a)).
Note that, even for a locally symmetrized state, the maximum position $n_\nu$ is still a stochastic variable, determined by the details of the random potential in (one half of) $\DD_d \owns n_\nu$ for a given disorder configuration.
In fact, the possible localization positions for given potential and eigenenergy can be found via the chain's ``localization landscape'', as outlined in \cref{app:localScattering}.
Distinct peaks at $n_\nu$ and $\P_{\DD_d}(n_\nu)$ occur when $\xi^\nu_d \gg \ell(\epsilon_\nu;w)$, that is, when the fictitious tunneling barrier is sufficiently strong and/or wide.
It may also often happen, however, that $\xi^\nu_d \sim \ell$, in which case the two fictitious wells practically merge into one, supporting a state peaked around the center $c_d$ of $\DD_d$.

Note that the localization length $\ell = \ell(\epsilon_\nu;w)$, indirectly determining each barrier strength via \cref{eq:fictitiousPotential}, generally depends in an involved manner on the energy and the disorder strength \cite{Rodriguez2006_JPAMG_39_14303_OnedimensionalModelsDisorderedQuantum}.
For the analysis carried out in the present work, it suffices to say that $\ell(\epsilon_\nu;w)$ overall decreases with increasing $w$ and $|\epsilon_\nu|$ (for fixed $w$).

\section{Eigenstate symmetrization statistics \label{app:eigenstateSymmetryStat}}  

The statistical distribution of the eigenstate IPR and CFS is analyzed in \cref{sec:statAnalysis} in terms of the relative occurrence of locally even/odd or left-/right-localized states induced by local symmetry.
To gain a more complete understanding of the localization and fragmentation for varying disorder strength $w$ and number of symmetry domains $D$, we here provide a statistical analysis of the symmetrization properties of the same eigenstates used for the IPR and CFS statistics in \cref{fig:ipr_cfs}.

To this end, we first label a given state $\ket{\phi^\nu}$ as ``domain-localized'' into a domain $\DD \equiv \DD^\nu$, if its total density $\sum_{n \in \DD^\nu}\rho_n$ within this domain exceeds a threshold value which we set to $0.95$.
For each such domain-localized state we define the domain ``density asymmetry''
\begin{equation}
 \delta_\nu = \left| \sum_{n \in \DD^\nu}\left[\rho_n - \rho_{\P_{\DD^\nu}(n)} \right]\right|
\end{equation}
with respect to local reflection $\P_{\DD^\nu}$ in this domain.
For a given LRD chain, we then define $N_{lr}$ as the number of eigenstates which have $\delta_\nu \geqslant \delta_{\text{thr}} \equiv 0.95$ (which are thus approximately confined to the left or right half of $\DD^\nu$), and as $N_{eo}$ the number of eigenstates with $\delta_\nu < 1 - \delta_{\text{thr}}$ (which are approximately even or odd under $\P_{\DD^\nu}$).
The remaining $N - N_{eo} - N_{lr}$ eigenstates are those which either do not localize to $95\%$ in a single symmetry domain, or do but do not reach $95\%$ density symmetry or asymmetry.

Note that $N_{eo}$ counts the even-odd \textit{pair} states described in \cref{sec:symmetrization}, but also single locally even or odd states with no quasidegenerate partner.
Similarly, $N_{lr}$ counts the left-right \textit{pair} states, but also single left- or right-localized states with no quasidegenerate partner.
Further, we point out that the above (global) threshold values are here simply chosen empirically to describe the present localization behavior; they could be relaxed to include less localized or non-symmetrized states in $N_{eo}$ and $N_{lr}$, or in principle also be refined to depend, e.\,g., on $w$.

\begin{figure*}[t!] 
\includegraphics[width=\textwidth]{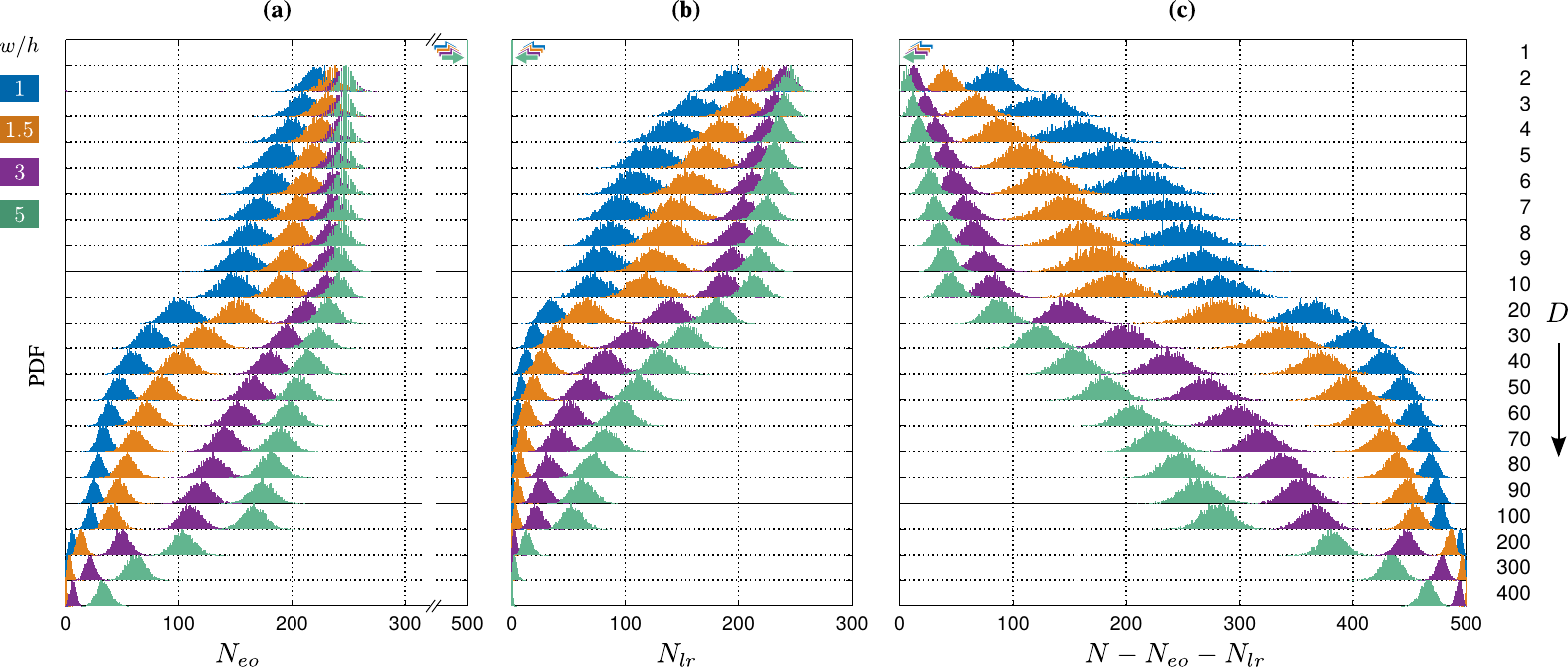}
\caption{
PDFs of the number of approximate
\textbf{(a)} locally even or odd eigenstates $N_{eo}$,  
\textbf{(b)} locally left- or right-localized eigenstates $N_{lr}$, and 
\textbf{(c)} remaining eigenstates $N - N_{eo} - N_{lr}$ (as defined in \cref{app:eigenstateSymmetryStat}) for varying $w$ and $D$, using the same ensemble of LRD chains as in \cref{fig:ipr_cfs}.
The colored arrows for $D = 1$ indicate a single bar at (a) $N_{eo} = 500$, (b) $N_{lr} = 0$, and (c) $N - N_{eo} - N_{lr} = 0$, for all $w$-values.
PDFs are normalized to maximum for each parameter combination.
}
\label{fig:lrsStateStat}
\end{figure*}

In \cref{fig:lrsStateStat} the PDFs (histograms) of $N_{eo}$, $N_{lr}$, and the remainder $N - N_{eo} - N_{lr}$ are shown for the LRD chain ensembles used in \cref{fig:ipr_cfs}, with varying disorder strength $w$ and number of symmetry domains $D \geqslant 1$.
For globally symmetric chains ($D = 1$) all states are even or odd, as evident from the single bars at $N_{eo} = N = 500$ and $N_{lr} = 0$ at any $w$.
For locally symmetric chains ($D > 1$), $N_{eo}$ and $N_{lr}$ are in general larger at stronger disorder, since the states are then more short-range-localized and thus less prone to leakage through domain interfaces.
For increasing $D$, we see that both $N_{eo}$ and $N_{lr}$ decrease at any $w$, since the domains become smaller on average, making interface localization as well as extension across domains more probable.
We also notice, however, that $N_{lr}$ decreases faster with $D$ than $N_{eo}$.
This predominance of locally even/odd states (which are overall more spatially fragmented) over left/right-localized states is in accordance with the CFS-distribution peaks at lower $\bar{f}$-values in \cref{fig:ipr_cfs}.

\section{IPR and CFS for random versus globally symmetric chains \label{app:noVersusGlobalSymmetry}}  

The downward jump of the mean IPR and, much more drastically, mean CFS distributions in \cref{fig:ipr_cfs} when switching on global symmetry (from $D = 0$ to $D = 1$) is here detailed in terms of individual $r_\nu$ and $f_\nu$ eigenstate profiles before averaging, and further analyzed for larger chain sizes $N$.

\begin{figure*}[t!] 
\includegraphics[width=\textwidth]{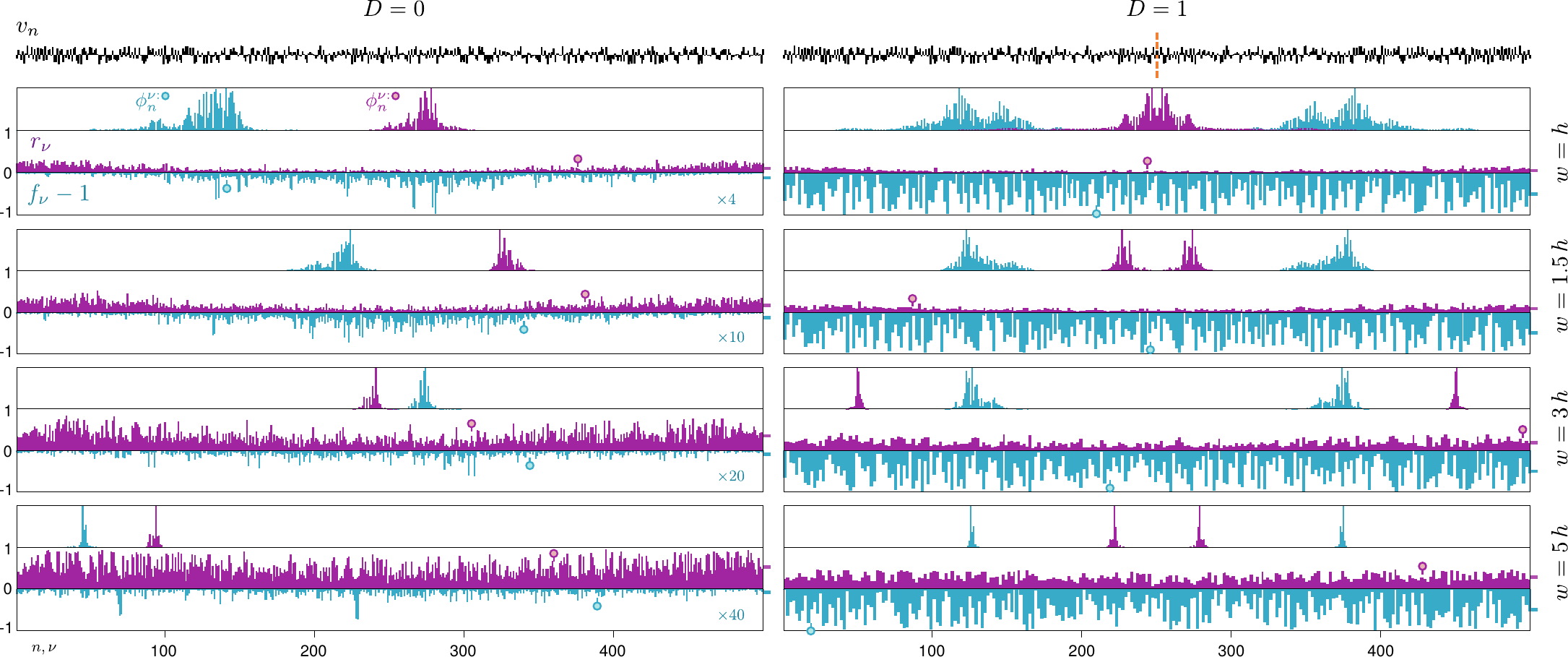}
\caption{
IPRs $r_\nu$ and CFSs $f_\nu$ of eigenstates $\ket{\phi^\nu}$, sorted in eigenenergies $\epsilon_\nu$, of \textit{(left column)} a random and \textit{(right column)} a globally reflection-symmetric disordered chain of size $N = 500$, with \textit{(top to bottom)} varying disorder strength $w$ applied to the same potential configuration $v_n$ shown in the top insets.
The right setup ($D = 1$) is the left setup ($D = 0$) symmetrized; the dashed vertical line indicates the symmetry axis.
The difference of $f_\nu$ from $1$ is plotted (magnified in left column by the indicated factors) for better visibility.
In each panel, the purple (blue) circle marks the state $\phi^\nu_n$ shown in the top row, corresponding to the $r_\nu$ ($f_\nu$) closest to the mean $\bar{r}$ ($\bar{f}$) indicated by colored horizontal bars on the right.
State magnitudes normalized to maximum are plotted.
}
\label{fig:ipr_cfs_N500_varDis_example}
\end{figure*}

Specifically, we consider in \cref{fig:ipr_cfs_N500_varDis_example} a random chain (left column, $D = 0$) and its symmetrized version (right column, $D = 1$), for which we present the complete distribution of $r_\nu$ and $f_\nu$ of all eigenstates $\ket{\phi^\nu}$, for the disorder strengths also used in \cref{sec:statAnalysis} (top to bottom panels).
Since the $f_\nu$ are relatively close to unity in the $D = 0$ case, we plot the difference $f_\nu - 1$ (further magnified by global factors for $D = 0$) to increase visibility.
Additionally, for each $(w,D)$-combination we show those eigenstates $\phi^\nu$ corresponding to the $r_\nu$ and $f_\nu$ closest to the mean values $\bar{r}$ and $\bar{f}$ (indicated by the small horizontal bars on the right of each panel), respectively.
They provide representatives of the kind of spatial profiles contributing to the statistical $\bar{r}$- and $\bar{f}$-distribution peaks. 

As expected, for the random chain ($D = 0$) the states simply become more localized when increasing $w$ (indicated also from the representative states shown), such that also the $r_\nu$ and $f_\nu$ overall increase (the $f_\nu$ get closer to $1$).
Further, states closer to the middle of the energy range for each setup are generally more extended, especially for weak disorder, as evident from the smaller $r_\nu$- and $f_\nu$-values there.

When turning on global symmetry ($D = 1$), all states have reflection-symmetric density, and especially the $f_\nu$-distributions change drastically:
They are shifted to much smaller values overall, with their mean $\bar{f}$ close to $1/2$ for any $w$.
Indeed, the shown representative states have density peaks symmetrically located about halfway from the center to the end of the chain (i.e., spacing $\xi \approx N/2$), yielding $f_\nu \approx 1/2$ (recall the description in \cref{sec:IPRvsCFS}).
Also, the fluctuations of the $f_\nu$ around $\bar{f}$ are rather homogeneous along the energy range, in contrast to the $D = 0$ cases; there is no suppression of the $f_\nu$ around the center, even for small $w$.
This occurs because the CFS depends primarily on the spacing of the density peaks, which can vary similarly for different $\epsilon_\nu$ as well as $w$, and rather less on the localization around the peaks themselves.
These features are not captured by the IPR, which is merely sensitive to the total site participation of a given density profile:
Since each state is globally symmetrized for $D = 1$, it contributes spatially by a double amount of sites, and the IPRs are approximately halved compared to the corresponding $D = 0$ setup (see also below).

\begin{figure}[t!] 
\center
\includegraphics[width=.7\columnwidth]{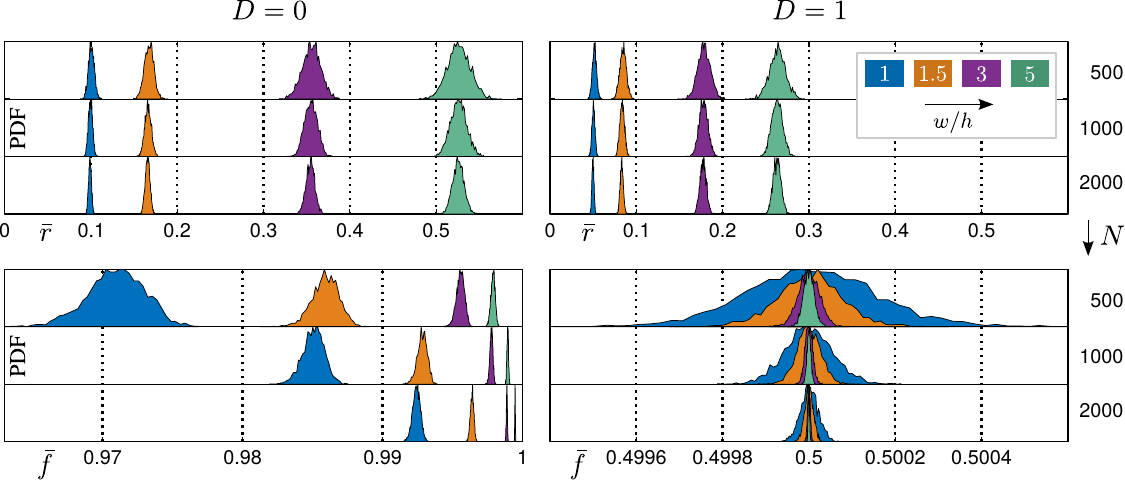}
\caption{
PDFs of the mean \textit{(top panels)} IPR $\bar{r}$ and \textit{(bottom panels)} CFS $\bar{f}$ for chains with \textit{(left panels)} no and \textit{(right panels)} global reflection symmetry, for different chain lengths $N$ and varying disorder strength $w$, using $3000$ configurations for each parameter combination.
}
\label{fig:ipr_cfs_varN}
\end{figure}

The above behavior of the state-resolved $r_\nu$ and $f_\nu$ are in accordance with the peak values observed in the IPR and CFS distributions for $D = 0,1$ in \cref{fig:ipr_cfs}, which we now show remain similar for longer chains.
Figure \ref{fig:ipr_cfs_varN} shows the PDF of the IPRs and CFSs for chains of length $N = 500, 1000, 2000$, with no symmetry and with global symmetry, for varying $w$.
For the IPR, we see that the peak $\bar{r}$-values remain practically the same when changing $N$ (since the spatial ``participation'' of the states does not change on average) and are halved when turning on the symmetry (whence the participation is doubled).
For the CFS, we firstly observe that the peak $\bar{f}$-values jump to $1/2$ when imposing global symmetry, as explained above, for all chain lengths $N$.
On the other hand, for $D = 0$ the PDFs are shifted to larger $\bar{f}$.
This is because the CFS depends on the relative extent of a state to the total chain size:
Its cumulative summand (see \cref{eq:cfs}) increases when the portion of sites with vanishing density increases.
For all cases, we also see that the overall statistical fluctuations of the $\bar{r}$- and $\bar{f}$-distributions around their peak values decrease with increasing $N$.

Finally, a technical remark is in order.
Computing the statistical IPR and CFS distributions (here for $3000$ configurations) of the longer chains ($N = 1000,2000$) would be prohibitive computationally at the precision necessary to guarantee the definite parity of all eigenstates in the $D = 1$ case \cite{MpackGMP}.
Therefore, we have here exploited the global symmetry to bring the Hamiltonian $H$ into a block-diagonal form with the same spectrum, with the parity-definite eigenvectors obtained from those of the blocks.

Specifically, for $D = 1$, $H$ is a real centrosymmetric tridiagonal matrix which, following the procedure described in e.\,g. Ref.\,\cite{Rontgen2018_PRB_97_35161_CompactLocalizedStatesFlat}, can be block-decomposed into the two matrices
\begin{equation}
H^{\pm} =
\begin{bmatrix}
v_{1}  & h  &        &        &        &  \\
h  & v_{2}  &      h &        &        &  \\
   & \ddots & \ddots & \ddots &        &  \\
   &        &        &  \hspace{-5em} h  & \hspace{-2em} v_{N/2-1} & \hspace{-2em} h \\
   &        &        &        & \hspace{-5em} h    & \hspace{-2em} v_{N/2} \pm h
\end{bmatrix}
\end{equation}
of halved size for even $N$.
If $x^{\mu} = [x_{1}^{\mu},\ldots{},x_{N/2}^{\mu}]^{\top}$ ($\mu = 1,\ldots{},N/2$) is an eigenvector of $H^{\pm}$ with eigenvalue $\epsilon_{\pm}^{\mu}$, then
\begin{equation}
	[x_{1}^{\mu},\ldots{},x_{N/2}^{\mu},\pm x_{N/2}^{\mu},\ldots{},\pm x_{1}^{\mu}]^{\top}
\end{equation}
is an eigenvector of $H$ with the same eigenvalue.
Thus, all eigenvalues and associated (even or odd) eigenvectors of $H$ are obtained, while the matrices $H^\pm$ can be diagonalized using standard double precision arithmetic (with, e.\,g., the \texttt{eig} function in MATLAB$^\circledR$).

Note that a decomposition of $H$ like above is not applicable to \textit{locally} reflection-symmetric chains, where the corresponding local site permutations do not commute with $H$.
Thus, computing eigenstates for $D > 1$ would require ever increasing arithmetic precision for increasing $N$, especially for large $w$.
We stress, however, that the purpose of the present study is \textit{not} to capture the effect of local symmetries in the large $N$ limit, but the dependence on the number of symmetry domains $D$ in \textit{finite} chains.
The size $N = 500$ of the LRD chains is simply chosen large enough (a) to be able to vary $D$ of a substantial range and (b) to extract a meaningful statistics (e.\,g. the mean $\bar{r}$ and $\bar{f}$) from individual chains.
In fact, in the $N \to \infty$ limit any (finite) local symmetry domain would be of negligible relative size, and thus adjacent domains would simply behave as random consecutive scatterers, each with some internal structure and multiple resonant levels.

\section{Eigenenergy spectra of LRD chains \label{app:spectralStat}}

In \cref{fig:setupExample} of the main text, the scale of the energy axis is too small to discern the effect of the local symmetries of the chain on its spectrum, which we briefly comment on here.

\begin{figure}[] 
\center
\includegraphics[width=.6\columnwidth]{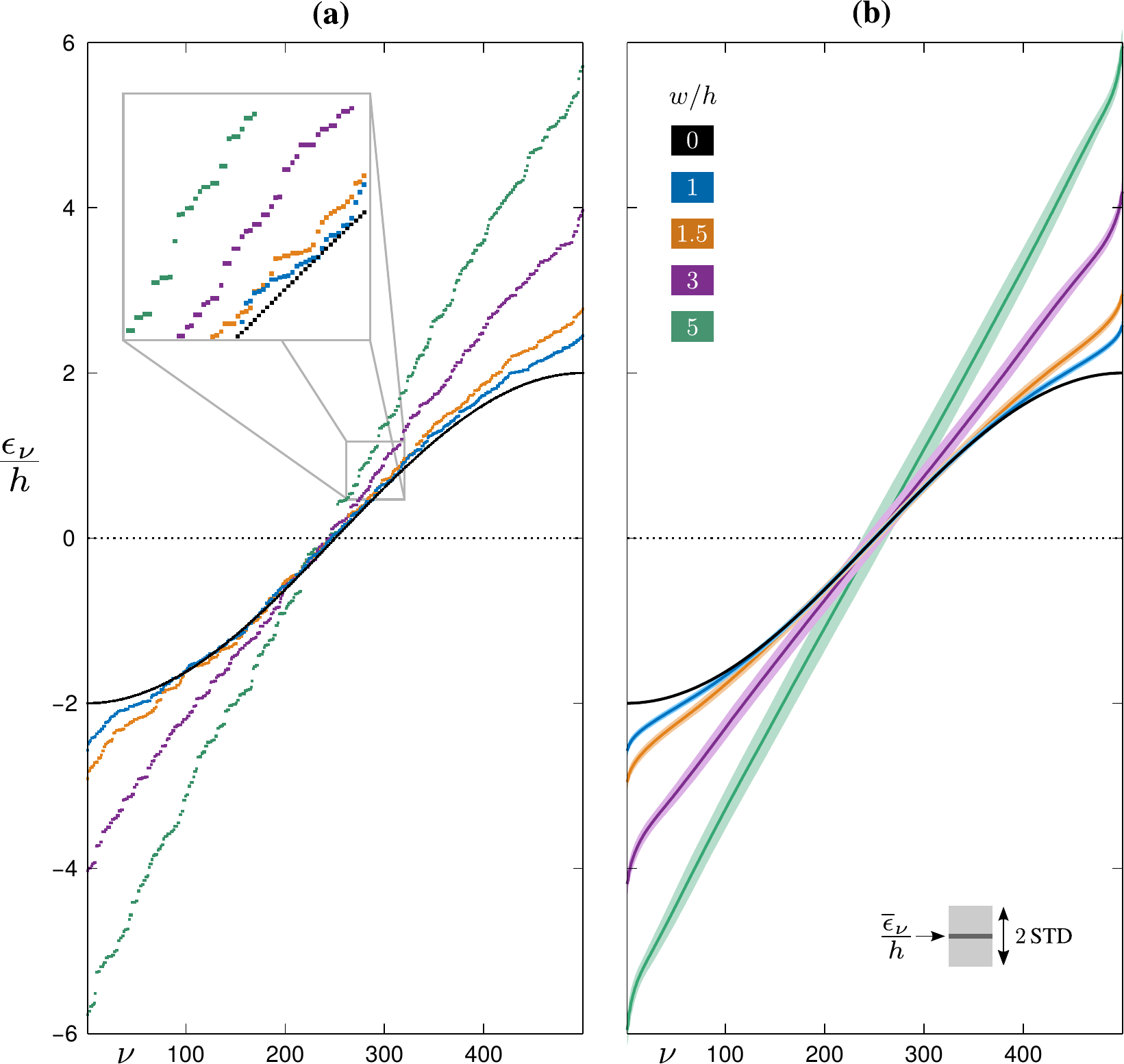}
\caption{
\textbf{(a)} Eigenenergy spectra $\epsilon_\nu$ of the LRD chain configuration of \cref{fig:setupExample} (with $D = 10$ symmetry domains) for varying disorder strength $w$, and \textbf{(b)} ensemble average (solid lines) of 3000 LRD configurations with the same $D,w$ as in (a), plotted on top of shaded stripes of width equal to two standard deviations at each $\nu$.
The inset in (a) reveals details of the single spectrum including multiple quasidegeneracies for larger $w$.
}
\label{fig:spectrum}
\end{figure}

In fact, the spectrum of an LRD chain closely resembles that of an uncorrelated random chain:
For sufficiently strong disorder, the eigenstates are so spatially localized that their eigenenergies (with contributions from the onsite potentials where the state magnitude is non-vanishing) are also uncorrelated; 
there is no so-called ``level repulsion'' between states.
Thus, the $\epsilon_\nu$ follow a roughly linear trend in the bulk of the spectrum, possibly with some additional low- or high-energy states at the ends of the spectrum (corresponding to states which happen to localize strongly on very low or high onsite potentials, respectively).

The only difference when imposing local reflection symmetries on a random chain is that multiple $eo$ and $lr$ state pairs may form, as explained in \cref{sec:symmetrization}, leading to corresponding quasidegeneracies in the spectrum.
Those can be distinguished in the inset of \cref{fig:spectrum}\,(a) showing the spectra of the LRD chain of \cref{fig:setupExample} for different disorder strengths (the $w = 3h$ spectrum is that of \cref{fig:setupExample}).
These quasidegeneracies do not affect the linear trend of $\epsilon_\nu$ on larger scale for large $w$.
They rather lead to a stronger fluctuation around the average spectrum $\bar{\epsilon}_\nu$ of the ensemble (of equivalent configurations) shown in \cref{fig:spectrum}\,(b).
Indeed, the standard deviation is measured to be a global factor $\approx 1.38$ larger than that for random chains without symmetries (not shown) for all considered $w$.

With decreasing $w$, the eigenstates become less localized, quasidegeneracies between $eo$-pair states and $lr$-pair states are gradually lifted (see inset in \cref{fig:spectrum}\,(a)).
The spectra approach the unperturbed spectrum of a homogeneous chain (shown in black in \cref{fig:spectrum}).
This is manifest more clearly in the ensemble average in \cref{fig:spectrum}\,(b).
There, we also see that the standard deviation around $\bar{\epsilon}_\nu$ naturally decreases towards the limit of the clean chain.

Finally, a comment is in order regarding the level spacing statistics of the considered model, as such a treatment is commonly used to characterize properties of disordered systems.
It is clear from the above that the level statistics will overall follow that of a chain with uncorrelated disorder, though with an increased contribution to spacings very close to zero at sufficient disorder strength, due to the quasidegeneracies induced by the local reflection symmetries.
Specifically, for relatively strong disorder the level spacings $\Delta\epsilon_\nu$ qualitatively follow a Poisson-like distribution with an additional enhancement for $\Delta\epsilon_\nu \to 0$.
When the disorder strength is lowered, a transition to Wigner-Dyson-like level statistics occurs due to level repulsion \cite{Haake2010____QuantumSignaturesChaos}.
Also the symmetry-induced quasidegeneracies are then increasingly lifted, since those levels repel each other too, thus depleting the contribution at $\Delta\epsilon_\nu \to 0$.
For all disorder strength values $w$ considered in the present context, however, the distributions remain practically Poisson-like, even for the weakest disorder with $w = h$.
We therefore refrain from presenting a study of level spacings for the model, which would additionally require a significant variation of parameters as well as larger chain sizes to be conclusive.
We emphasize, however, that the purpose here is to analyze the impact of local reflection symmetries on the eigenstates and transfer behavior of an explicitly finite chain.
A targeted statistical spectral analysis of the long chain limit is left as an interesting direction for future work.


%

\end{document}